\documentclass[twocolumn]{openjournal}

\usepackage[utf8]{inputenc}
\usepackage[T1]{fontenc}
\usepackage{float}
\usepackage{xcolor}
\usepackage{ulem}
\usepackage{graphicx}	% Including figure files
\usepackage{amsmath}	% Advanced maths commands
\usepackage{comment}
\usepackage{color}
\usepackage{hyperref}
\hypersetup{colorlinks=true,
	urlcolor=blue,  % color of external links
	linkcolor=blue,  % color of toc, list of figs etc.
	citecolor=blue, % color of links to bibliography
    menucolor=blue, % color for Acrobat menu buttons
    urlcolor=blue}  % color for \url{...} links
\newcommand{\ergs}{\ensuremath{\,\mathrm{erg}\,\mathrm{s}^{-1}}}

\newcommand{\hmpc}{\ensuremath{\;h^{-1}\mathrm{Mpc}}}
\newcommand{\chmpc}{\ensuremath{\;h^{-1}\mathrm{cMpc}}}

\newcommand{\phmpc}{\ensuremath{\;h^{-1}\mathrm{pMpc}}}

\newcommand{\chgpc}{\ensuremath{\;h^{-1}\mathrm{cGpc}}}

\newcommand{\invchmpc}{\ensuremath{\;h\,\mathrm{cMpc}^{-1}}}

\newcommand{\hcgpccc}{\ensuremath{\;h^{3}\mathrm{cGpc}^{-3}}}
\newcommand{\hkpc}{\ensuremath{h^{-1}\,\mathrm{kpc}}}
\newcommand{\kms}{\ensuremath{\,\mathrm{km}\,\mathrm{s}^{-1}}}

\newcommand{\hmsun}{\ensuremath{h^{-1}M_{\odot}}}
\newcommand{\msun}{\ensuremath{\,M_{\odot}}}

\newcommand{\lya}{\ensuremath{\mathrm{Ly}\alpha}}

\newcommand{\simlt}{\ensuremath{\lower.5ex\hbox{\(\,\buildrel<\over\sim\,\)}}}
\newcommand{\simgt}{\ensuremath{\lower.5ex\hbox{\(\,\buildrel>\over\sim\,\)}}}

\newcommand{\rhobar}{\ensuremath{\bar{\rho}}}

\newcommand{\astrid}{{\tt Astrid}}

\newcommand{\uchuu}{{\tt Uchuu}}
\begin{document}

\title{
Large-scale surveys of the quasar proximity effect
}

\author{
Rupert A.C. Croft$^{1,*}$, Patrick Shaw$^{1}$, Ann-Marsha Alexis$^{1}$, Nianyi Chen$^{2}$,  Yihao Zhou$^{1}$, Tiziana Di Matteo$^{1}$, Simeon Bird$^{3}$, Patrick Lachance$^{1}$ and Yueying Ni$^{4}$
}
\thanks{$^*$E-mail:rcroft@cmu.edu}
% List of institutions
\affiliation{$^{1}$ McWilliams Center for Cosmology, Department of Physics, Carnegie Mellon University, Pittsburgh, PA 15213 USA \\}
\affiliation{$^{2}$ Institute for Advanced Study, 1 Einstein Dr, Princeton, NJ 08540\\}
\affiliation{$^{3}$ Department of Physics \& Astronomy, University of California, Riverside, 900 University Ave., Riverside, CA 92521, USA\\}
\affiliation{$^{4}$ Harvard-Smithsonian Center for Astrophysics, Harvard University, 60 Garden Street, Cambridge, MA 02138, USA\\}

\begin{abstract}

The UV radiation from high redshift quasars causes a local deficit in the neutral hydrogen absorption (Lyman-$\alpha$ forest) in their spectra, known as the proximity effect. Measurements from small samples of tens to hundreds of quasars have been used to constrain the global intensity of the UV background radiation, but so far the power of large-scale surveys such as the Sloan Digital Sky Survey and the Dark Energy Spectroscopic Instrument (DESI) survey has not been used to investigate the UV background in more detail. We develop a CDM-based  halo model of the quasar proximity effect, which accounts by construction for the fact that quasars reside in overdense regions. We test this model on quasar Lyman-$\alpha$ spectra from the \astrid\ cosmological hydrodynamic simulation, which includes self-consistent formation of quasar black holes and the intergalactic medium surrounding them. Fitting the model to individual quasar spectra, we 
constrain two parameters, $r_{\rm eq}$ (the radius at which the local quasar radiation intensity equals the background), and the quasar bias $b_{\rm q}$ (related to host halo mass). We find that $r_{\rm eq}$ can be recovered in an unbiased fashion with a statistical uncertainty of
$25-50\%$ from a single quasar spectrum. Applying such fitting to samples of millions of spectra from the DESI (for example) would allow measurement of the UV background intensity and its evolution with redshift with high precision. We use another, larger-scale, lower resolution simulation (\uchuu) to test how such a large sample of proximity effect measurements could be used to probe the spatial fluctuations in the intergalactic radiation field. We find that the large-scale structure of the UV radiation intensity could be mapped and its power spectrum measured on  $100\sim 1000$ h$^{-1}$Mpc scales. This could allow the large-scale radiation field to join the density field as a dataset  for constraining cosmology and the sources of radiation. 

\end{abstract}


\section{Introduction}

The proximity effect is the observed reduction of Ly$\alpha$ absorption in the intergalactic medium (IGM) close to quasars, primarily driven by enhanced photoionization from the quasar itself, which locally elevates the ionization rate of neutral hydrogen \citep{carswell82, bajtlik88,lu91}. Early studies of this phenomenon were limited to small samples of quasars, with significant statistical progress made in the 1990s and 2000s using samples on the order of tens to hundreds of quasars \citep[e.g.,][]{bechtold94, scott2000}. However, modern large-scale spectroscopic surveys, such as the Baryon Oscillation Spectroscopic Survey (BOSS; \citealt{dawson13}), now provide samples of hundreds of thousands of quasars, with the Dark Energy Spectroscopic Instrument (DESI; \citealt{desi16,desi25}) delivering millions in the near future. This explosion in data enables us to revisit the quasar proximity effect with unprecedented statistical power, opening new avenues for analysis and discovery.

The proximity effect offers an indirect probe of quasar luminosity, which can be used to estimate the photoionization rate in the surrounding IGM. Ly$\alpha$ absorption in the IGM depends sensitively on the neutral hydrogen density \citep{gunn65} and the overall ionizing background, comprising contributions from the ultraviolet background (UVBG) and the quasar’s own radiation field. Historically, studies of the proximity effect have been used to estimate the UVBG, by assuming the quasar luminosity and subtracting its contribution to the local ionization rate \citep[e.g.,][]{scott2000, calverley11, dall08}. Conversely, given assumptions about the UVBG, the proximity effect can provide insights into the IGM density distribution (e.g., \citealt{jalan21},\citealt{rollinde05},\citealt{guimar07},\citealt{dodorico08}), potentially allowing estimates of the mass of dark matter halos hosting the quasars (e.g., \citealt{faucher08,kim04}).

One possible application of large-scale surveys is using the proximity effect as a cosmological tool. By comparing the quasar luminosity inferred from proximity effect measurements to its observed luminosity, one could, in principle, determine the quasar’s luminosity distance and thus probe cosmological parameters \citep[e.g.,][]{phillipps02}. While early work  encountered challenges due to uncertainties in quasar lifetimes and variability, the substantial increase in data from upcoming surveys, combined with improved models of quasar fueling and light curves, could allow for meaningful progress.

Large scale fluctuations (\citealt{zuo92,meiksin03,gontcho14}) in the intergalactic
radiation field are generated by the clustered sources of radiation. These radiation fluctuations can modulate observables
such as the \lya\ forest on large scales, offering ways to study them. Recently, \cite{longhirata23} showed how bias and shot-noise parameters describing the sources can be constrained by observations that cross-correlate galaxy surveys and the \lya\ forest. The proximity effect due to quasars can also have a role, allowing us to map out the radiation field, albeit sparsely, and we will explore this aspect also.

To explore these opportunities, we will leverage the \astrid\ hydrodynamical simulation \citep{bird22}, which provides not only a detailed model of the IGM but also includes realistic quasar populations with self-consistent light curves. 
\astrid’s large volume allows us to model rare, massive quasars while simultaneously capturing the smaller halos and their interactions with the IGM. This makes it an excellent tool for testing theoretical models of the quasar proximity effect, with the aim of refining predictions for future observational measurements.

Our plan for the paper is as follows: in Section \ref{halo} we describe our model for the proximity effect, based on the clustering of dark matter halos and matter in a CDM universe. We simulate the quasars and the proximity effect using the \astrid\ cosmological hydrodynamic simulation in Section \ref{astridsec}. In Section \ref{anasim} we bring the two together, testing the halo model by analyzing the outputs of the simulation. In Section \ref{largefield} we use another larger 
 simulation to discover how the large scale cosmic radiation intensity can be mapped out using the proximity effect, and in Section \ref{sumdisc}, we summarize and discuss what we have learned and outline directions for future work.

%\cite{dallaglio10} analyzed prox effect in hydro sims

\section{A halo model for the quasar proximity effect}
\label{halo}

When modeling the proximity effect, we aim to produce a theoretical profile of the transmitted \lya\ forest
flux in a quasar spectrum as a function of distance from the quasar. We will use this model to analyse both the mean \lya\ profile
around many quasars and also the \lya\ profile
for individual objects. We restrict ourselves in this paper
to tests of the model at $z=3$, but we expect that the model should be applicable to the \lya\ forest from redshifts $z=2\sim 4$.

The environment around quasars is known to be 
overdense with respect to the cosmic mean (e.g., \citealt{jalan21}), and 
many previous papers have addressed how this will
bias measurements of the UVBG intensity made from it.
The work of \cite{faucher08}  (see also \citealt{kim04}) made this a central
point of their analysis, showing how the \lya\ forest 
in the proximity effect region could be analysed
in order to constrain the host halo mass of a quasar. In order to include this effect in our analysis, we will assume that quasars exist in dark matter halos, which trace the large-scale structure expected in 
CDM cosmological models. This combination of a dark matter halo density profile (e.g., NFW, \citealt{nfw}) and the linear theory CDM clustering makes up the so-called halo model 
 (\citealt{cooray02,seljak00,neyman52}). It has been applied successfully in many contexts, including galaxy clustering (e.g., \citealt{zheng07}), Lyman-limit
systems (\citealt{theuns24}), and weak lensing (\citealt{mandelbaum05}).

In the case of the \lya\ forest, we use the dark matter
halo profile and clustering as a starting point, assuming that the neutral hydrogen traces the dark matter structure (\citealt{bi91}), except for the local
ionizing effect of the quasar. 

\subsection{Dark matter profile}

The fit to dark matter clustering is inspired
by the fact that the clustering of galaxies and of dark matter
can be described (\citealt{cooray02})
by terms which are a combination of internal 
halo structure (the "one halo" term) and large-scale, linear clustering (the "two halo" term).
 The cross-correlation function of the quasar host halo centers (where the quasars are assumed to reside) and the dark matter density in this
 case is

\begin{equation}
\xi_{q\rho}(r) = \frac{A_{\rm NFW}\delta_{c}}{cr(1+cr)^{2}}
+b_{q}\xi_{\rm CDM}(r),
\label{eqnfw}
\end{equation}
where $\delta_{c} = ((200/3)c^{3})/
       (\ln(1+c)-(c/(1+c))$.
Here we assume that the internal halo structure follows
the \cite{nfw} relation (the concentration $c$ and 
amplitude A$_{\rm NFW}$ are
free parameters) and 
$\xi_{\rm CDM}$ is the linear theory Cold Dark Matter correlation function
(computed using \citealt{lewis11}). $b_{\rm q}$ is a linear bias parameter relating
quasar and matter clustering on large scales.

In Section \ref{astridsec} below we will use a cosmological hydrodynamic simulation of quasars
to investigate the terms in Equation \ref{eqnfw}
and the parameters $b_{q}$, $c$ and $A_{\rm NFW}$.

\subsection{From dark matter to \lya\ optical depth}
Equation \ref{eqnfw} represents an averaged dark matter density profile around a quasar. In order to predict 
the \lya\ forest due to neutral hydrogen around quasars, we  use the Fluctuating Gunn-Peterson Effect (\citealt{weinberg97,croft97}):
\begin{equation}
\tau=A \rho^{(2-0.7\gamma)}T^{-0.7}\Gamma^{-1}
\label{fgpa}
\end{equation}
where $\tau$ is \lya\ optical depth, and $\rho$ is the dark matter density
in units of the cosmic mean. The temperature $T$ is related to the density by 
a power-law relationship $T\propto\rho^\gamma$ (\citealt{hui97}). We assume  $\gamma=0$, as favored by observations at redshift $z\sim 3$ (e.g., \citealt{bolton08}), but our results are not sensitive to the choice of $\gamma$. In Equation \ref{fgpa}, $\Gamma$ is the photoionization rate of hydrogen, proportional to the intensity of the radiation field in the optically thin approximation, so that $\tau$ is inversely proportional to it. The 
parameter $A$ is an overall normalizing parameter that we set by requiring the mean
transmitted flux $F=e^{-\tau}$ in the \lya\ forest to be equal to values measured in 
observational data. We use the values of \cite{kim07}, so that $\langle F \rangle=0.696$
at $z=3$.

Because $\tau$, and also the observed quantity, $F$  are non-linear functions of $\rho$, if we want to predict for
example the mean profile of $\tau$ or of $F$ around a quasar it is not enough to use the mean $\rho$ profile (from Equation \ref{eqnfw}) - we
also need the probability density function, pdf, of $\rho$ at each point in space. 

\begin{figure}
    \centering
    \includegraphics[width=\linewidth]{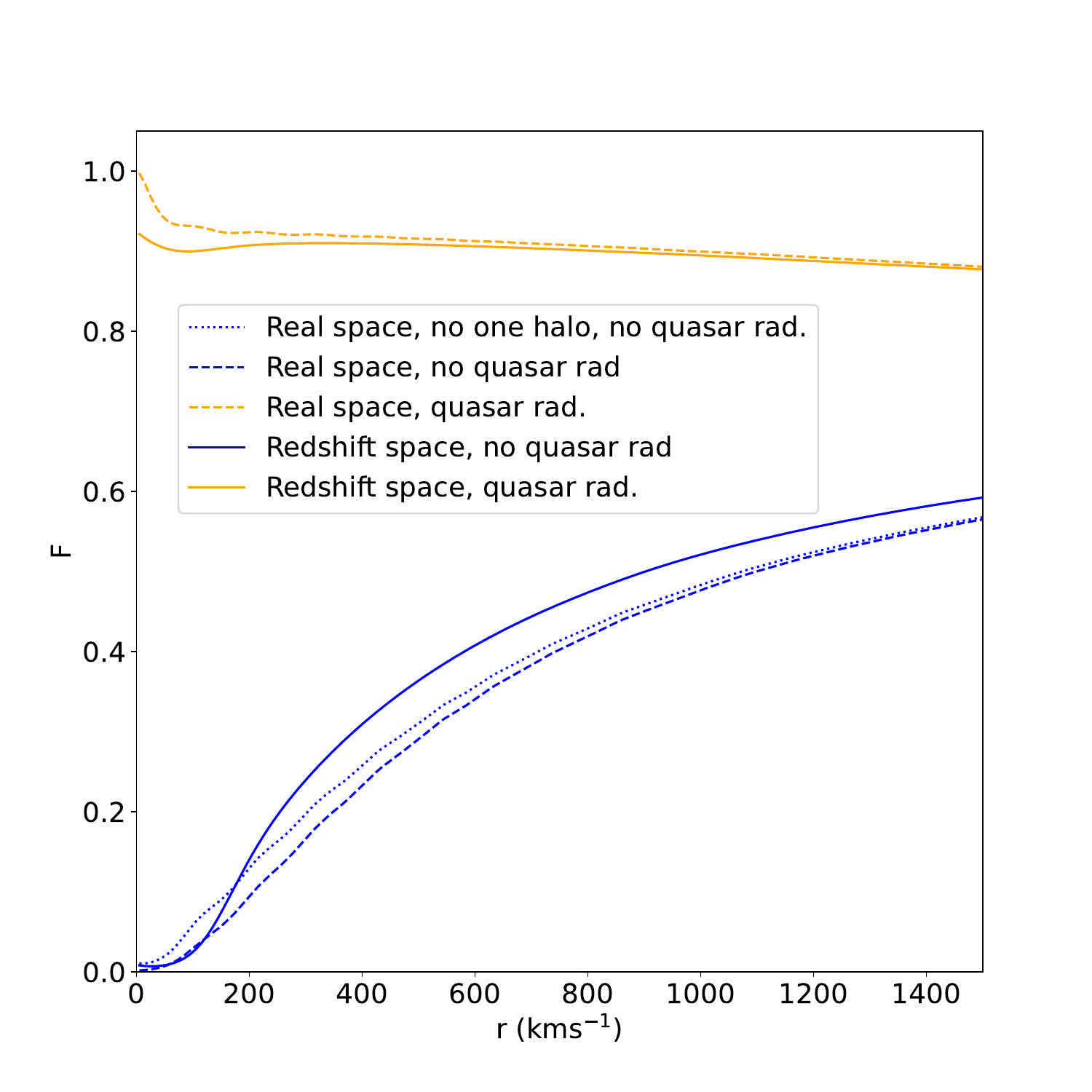}
    \caption{The halo model
    of \lya\ transmitted flux as a function of separation from quasars. We show the model
    curves without the effect of local ionizing radiation as blue lines and with the
    quasar radiation as yellow lines. We show the model in real space and redshift space, and also (in real space) without the one-halo term (Equation \ref{eqnfw}).}
    \label{haloexample}
\end{figure}

We follow the work of \cite{bi97} in assuming that the pdf of the density on physical scales relevant to the \lya\ forest (pressure smoothing scale of $\sim 100 \hkpc$, \citealt{peeples10}) can be modelled using a lognormal 
distribution (e.g., \citealt{coles91}). In
the lognormal distribution, given a mean density ${\bar{\rho}}$ and standard deviation
$\sigma$, the pdf (for a given radius $r$ from the quasar in our case) is given by
\begin{equation}
\label{eq:lognormal_pdf}
P(\rho \mid r) \;=\;
\frac{1}{\rho \,\sqrt{2\pi}\,\sigma(r)} 
\exp\!\Biggl[
  -\frac{\bigl(\ln \rho \;-\;{\bar{\rho}}(r)\bigr)^{2}}{2\,\sigma(r)^{2}}
\Biggr].
\end{equation}
In the case of our halo model, the mean density 
$\bar{\rho}$  varies as a function of distance from the quasar,  so that $\bar{\rho}(r)=1+\xi_{q\rho}(r)$,
and $\xi_{q\rho}$ is computed from Equation \ref{eqnfw}. We assume that $\sigma(r)=f \bar{\rho}$ where $f$ is a constant, i.e. the
rms variation of the density about the mean is a constant factor times the density. This is motivated by simulation data (Section \ref{astridsec}), from which we also take the value $f=1.25$ .  Recent work by \cite{ondro24} has shown that lognormal simulations can capture the thermal properties of the intergalactic medium quite well and be used
for cosmological inference from the \lya\ forest. We use Equation \ref{fgpa} to convert the $\rho$ values in Equation \ref{eq:lognormal_pdf} into optical depths, $\tau$, yielding the pdf of $\tau$ at different distances from the quasar,
$P(\tau\mid r)$.

\subsection{Quasar photoionization}

At this point in construction of the proximity effect model, we have a set of  \lya\ optical depth pdfs, each one corresponding to different distance from the quasar. We now introduce the local ionizing radiation from the quasar, using an inverse square law attentuated by an exponential attenuation
length:

\begin{equation}
\tau_{\rm prox}=\tau \left( \frac{r^{2}_{\rm eq}}{r^{2}} \right) e^{-r/r_{\rm att}}
\label{tauprox}
\end{equation}
Here the attenuation length is taken to be $r_{\rm att}=88 \phmpc$ at the redshift we concern ourselves with in this paper, $z=3$ (\citealt{theuns24}). In the rest of the paper we will use 
$\phmpc$ to refer to proper length units and $\chmpc$ for comoving length units. In the literature, the former are usually used for physical quantities related to the proximity effect, and the latter for large-scale structure measures, simulation box sizes and so on.

The quantity
$r_{\rm eq}$ is the radial distance from the quasar at which the local ionizing radiation from the 
quasar becomes equal to the overall ionizing background 
radiation:

\begin{equation}
    r_{\rm eq} = \sqrt{\frac{N_{\gamma} \sigma_{\rm HI}}{4\pi \Gamma_{\rm HI}}},
    \label{req_eqn}
\end{equation}

where $N_{\gamma}$ is the rate of ionizing photon emission from the quasar (in s$^{-1}$), $\sigma_{\rm HI}$ is the cross-section for photoionization of hydrogen (equal to $6.3 \times 10^{-18}$ cm$^{2}$), and $\Gamma_{\rm HI}$ is the photoionization rate due to the background radiation (in s$^{-1}$). Following our convention above, we refer to $r_{\rm eq}$ in proper $h^{-1}$Mpc  (\phmpc) throughout this paper.

\subsection{Redshift space}
The \lya\ optical depths in the model so far are in real space, but observations include the effects of peculiar velocities
and thermal broadening, being in  redshift space.
As in \cite{croft99}, 
we model the former using a spherical infall model
for large scale inflow (\citealt{yahil85}) and a small scale
random velocity dispersion (\citealt{davis83}). 
The inflow model is
\begin{equation}
v_{\rm infall}(r)= -\frac{1}{3}\Omega_{0}^{0.6}H_{0}r\frac{\delta(r)}{[1+
\delta(r)]^{0.25}}
\end{equation}
where $\delta(r)$ is the matter overdensity averaged within radius
$r$ of the quasar:
\begin{equation}
\delta(r)=\frac{3}{r^{3}}\int^{r}_{0}\xi_{q\rho}(x)x^{2}dx
\end{equation}

It was found by \cite{croft99} that this velocity field model is substantially better than linear theory
for the inflow pattern around massive halos. We expect this
to be true of quasar host galaxies, although as we see in Figure \ref{haloexample} the effects of redshift distortions
at $z=3$ are relatively small.
Because $v_{\rm infall}(r)$ is not expected to describe the virialized
regions of large halos, we follow \cite{croft99} and truncate
$v_{\rm infall}(r)$  on small scales by multiplying by an
exponential, $\exp{-(\delta/50)}$.

The random velocity dispersion we use is an exponential model,
so that the distribution function of velocities is
\begin{equation}
f(v)=\frac{1}{\sigma_{12}\sqrt{2}}\exp\left(-\frac{\sqrt{2}|v|}{\sigma_{12}}
\right).
\label{fv}
\end{equation}
here $\sigma_{12}$ is the pairwise velocity dispersion of quasar-neutral hydrogen atom
pairs, which we assume to be independent of 
pair separation. Based on 
simulation results we take this value to be $\sigma_{12}=100 \kms$. The thermal broadening contribution is also added at this stage, using a Gaussian approximation to a Voigt
profile with $\sigma_{\rm T}=\sqrt{k_{\rm B}T/m_{\rm H}}$.  
Because changing the value of $\sigma_{12}$ and $\sigma_{\rm T}$ only affect the inner one or two  pixels at DESI resolution,
the results are insensitive to changes in their values.

In order to convert our \lya\ optical depth pdfs in each pixel
into redshift space, we convolve with the velocity model
\begin{equation}
\label{eq:velocity_convolution}
\widetilde{P}(\tau \mid r)
\;=\;
\int_{0}^{\infty} \int_{0}^{\infty}
P(\tau' \mid r') \;
W(r' \to r)
\; d\tau'\,dr'.
\end{equation}

Here, \(W(r' \to r)\) represents the kernel describing how optical depths from real-space position \(r'\) map to the observed coordinate \(r\) due to the velocity field model, including infall and dispersion terms.

Once we have the velocity-shifted optical-depth distribution $(\widetilde{P}(\tau \mid r$)), we compute the mean transmitted flux as:
\begin{equation}
\label{eq:mean_flux_continuous}
\langle F \rangle(r) 
\;=\; 
\int_{0}^{\infty} 
\exp\bigl[-\tau\bigr]
\,\widetilde{P}(\tau \mid r)
\, d\tau.
\end{equation}

In Figure \ref{haloexample}, we show the mean transmitted flux profile around quasars for various steps in the model. The blue curves are the results without the local quasar ionizing radiation, where we can see that the absorption increases monotonically down to $r=0$, the position of the quasar. This effect of quasars inhabiting overdense regions is expected, and we also see that the effect
of the one-halo term (in Equation \ref{eqnfw}) is relatively small by comparing the dotted blue (which excludes it) and dashed blue line. Those lines are both in real space, and the effect of redshift space distortions can be seen in the solid blue line- the infall squeezes the absorption profile compared to that in  real space.  

The yellow lines in Figure \ref{haloexample} show
the model once local quasar radiation is included. In this example. the parameter $r_{\rm eq}=10 \phmpc$, equal to 
what is expected for a bright ($M_{\rm UV}=-28$) quasar.
The $x$-axis of the plot shows distances as far
as $1400 \kms$, which is  3.2 $\phmpc$ at $z=3$.
The proximity effect is dramatic on the scales
plotted, with flux in the quasar spectrum rising to
the unabsorbed level $F=1$.

\section{Cosmological hydrodynamic simulation of quasars and their proximity effect}

\label{astridsec}

With cosmological simulations of structure formation reaching the volumes required to include some bright quasars (e.g., \citealt{dimatteo17, springel18, duboi16}) we are able to test our halo model for the proximity effect with some detailed simulated data. 

\subsection{The Astrid simulation}

We use a redshift $z=3$ snapshot from the \astrid\ simulation for our study. \astrid\ is the largest (in terms of particle number and total compute hours) hydrodynamic simulation run to $z=0$ so far (\citealt{bird22,ni24,chen25}).  It has sufficient mass and spatial resolution to model the intergalactic medium that causes the \lya\ forest, as well as a large enough volume to sample the
quasar luminosity function (\citealt{ni22}).

\begin{figure*}[ht]
    \centering
    \includegraphics[width=\linewidth]{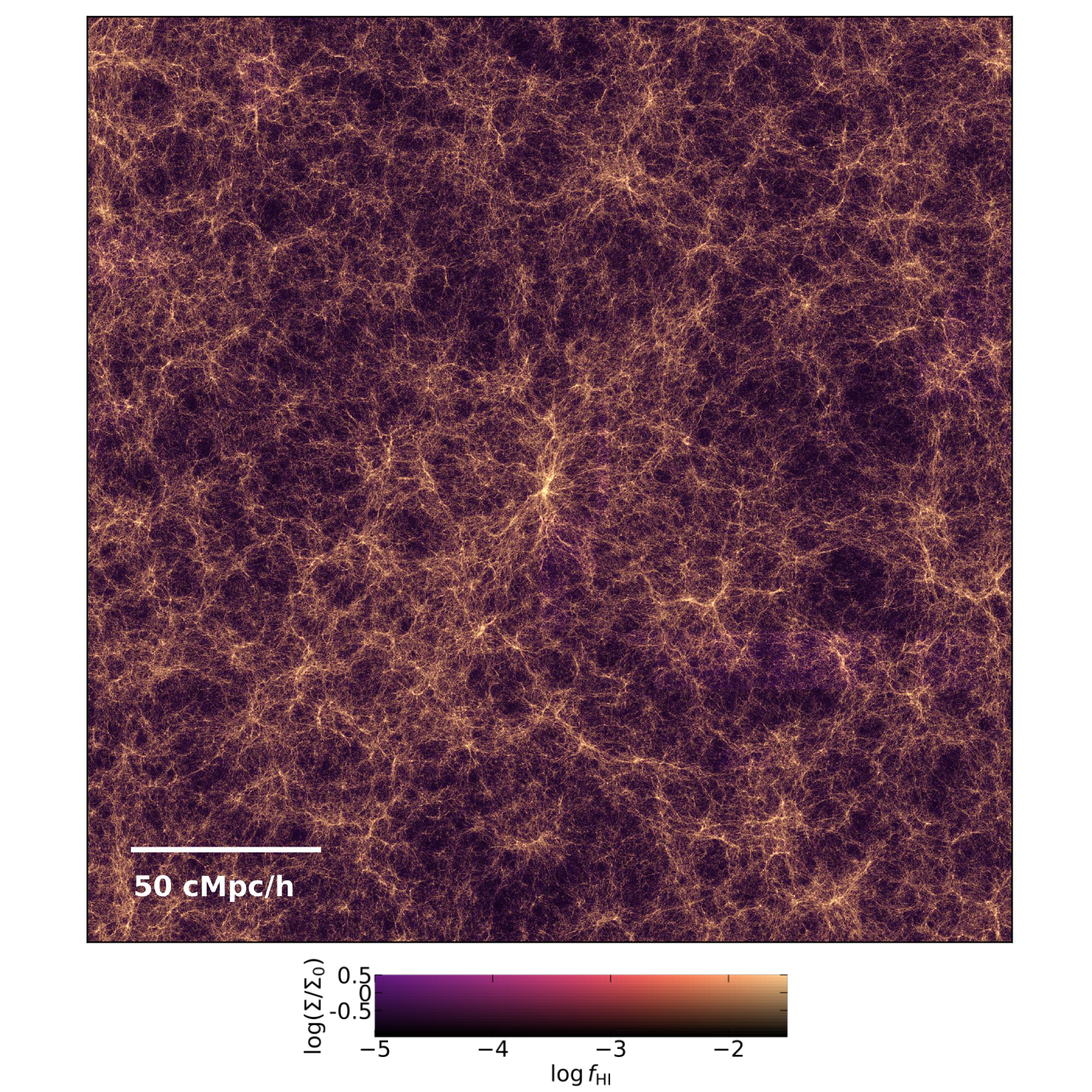}
    \caption{A slice through the \astrid\ simulation
    at $z=3$. The brightest quasar is
    in the center of the volume. We plot the gas
    density color-coded by neutral fraction and density (see color scale) in a slice of thickness 25 \chmpc. The UV radiation field used to compute the neutral fractions in this image is the uniform one from the simulation, without the localized effects of quasar radiation.  }
    \label{astridq}
\end{figure*}

\astrid\ was run using an updated version of the \texttt{MP-Gadget} code \citep{feng18}, a highly scalable variant of the \texttt{Gadget-3} cosmological structure formation code \citep{springel01,springel05}. It retains the core algorithms from \cite{feng16} but has undergone substantial improvements in speed and scalability.

 The simulation consists of $5500^3$ cold dark matter (DM) particles within a cubic volume of $250 \hmpc$ per side, along with an equal number of SPH hydrodynamic mass elements at the start. The initial conditions are set at redshift $z=99$, and the simulation snapshot we use is from  $z=3$. The adopted cosmological parameters follow \citep{planck18}, with values of $\Omega_0=0.3089$, $\Omega_\Lambda=0.6911$, $\Omega_{\rm b}=0.0486$, $\sigma_8=0.82$, $h=0.6774$, $A_s = 2.142 \times 10^{-9}$, and $n_s=0.9667$. The initial mass resolution is $M_{\rm DM} = 6.74 \times 10^6 \hmsun$ for dark matter and $M_{\rm gas} = 1.27 \times 10^6 \hmsun$ for gas, with a gravitational softening length of $\epsilon_{\rm g} = 1.5 \hkpc$ for both components.

 Below, we summarize key aspects of the hydrodynamical implementation and galaxy formation models of the version of \texttt{MP-Gadget} used to run \astrid, with a more detailed description available in \cite{bird22}. 

Gravitational interactions in \astrid\ are computed using the TreePM algorithm. The simulation employs the pressure-entropy formulation of smoothed particle hydrodynamics (pSPH) \citep{hopkins13,read10} to solve the Euler equations, while star formation and stellar feedback models closely follow those in \cite{feng16}. Gas cooling occurs both radiatively, following \cite{katz96}, and through metal-line cooling, where metallicity-dependent cooling rates are estimated by scaling a solar metallicity template \citep{vogelsberger14}.

To model patchy reionization, \astrid\ incorporates a spatially varying ultraviolet background using a semi-analytic approach informed by radiative transfer simulations \citep{battaglia13}. Within ionized regions, the ionizing background follows \cite{fg20}, while self-shielding is implemented following \cite{rahmati13}, ensuring gas remains neutral above a density threshold of $0.01$ atoms/cm$^{-3}$. At the redshifts of relevance here ($z=3$), the IGM is fully ionized.

The star formation model is based on the multi-phase prescription of \cite{springel03}, incorporating molecular hydrogen formation and its influence on star formation at low metallicities, as described by \cite{krumholz11}. Newly formed stars typically have a mass equal to one-quarter of the parent gas particle, resulting in $M_{\rm star} \sim 3\times 10^5 \hmsun$ in most cases. Supernova feedback follows \cite{okamoto10}, with Type II supernova-driven winds having speeds proportional to the local one-dimensional dark matter velocity dispersion. Wind particles remain hydrodynamically decoupled for a period, during which they do not contribute to pressure forces or accrete onto black holes.

The simulation also includes models for helium reionization and the impact of massive neutrinos. Metal enrichment from AGB stars, Type II supernovae, and Type Ia supernovae is incorporated following \cite{vogelsberger13} and  \cite{pillepich18}, though the implementation is distinct, using independently developed yield tables, as detailed in \cite{bird22}.

Supermassive black hole (SMBH) physics in \astrid\ is modeled similarly to \texttt{Bluetides} \citep{feng16}, based on earlier work by \cite{sdh05,dsh05}. However, in contrast to \texttt{Bluetides}, SMBH seeding follows a mass distribution rather than a fixed universal seed mass.  BH seeds are placed in halos that exceed both a halo mass threshold of $5 \times 10^{9} h^{-1} M_{\odot}$ and a stellar mass threshold of $2 \times 10^{6} h^{-1} M_{\odot}$. The seed mass $M_{\mathrm{sd}}$ is drawn randomly from a power-law distribution between $3 \times 10^{4} h^{-1} M_{\odot}$ and $3 \times 10^{5} h^{-1} M_{\odot}$ (there is no correlation between seed mass and halo mass, assuming a halo mass above the threshold). The simulation then tracks the evolution of BHs, modeling their growth primarily through gas accretion—calculated via a Bondi-Hoyle-like prescription—and mergers, which are governed by a dynamical friction model \citep{chen22} that accounts for interactions with surrounding stars and dark matter.
BH mergers occur only when two BHs become gravitationally bound based on their relative velocities and accelerations, typically resulting in mergers approximately $200\,\mathrm{Myr}$ after their initial encounter. AGN feedback is implemented by coupling 5\% of the BH accretion luminosity thermally to surrounding gas. This approach ensures realistic growth and merger histories for BHs within the simulated cosmological environment.

\subsection{Quasars in the Astrid simulation}

In Figure \ref{astridq} we present an
illustrative image of a slice through
the gas density of \astrid. The panel is centered on the position of the 
most luminous quasar at $z=3$, which has a bolometric luminosity of $2.5\times10^{47}$\ergs and
is hosted by a halo of mass $7 \times 10^{13}$\msun. The black hole mass is $1.2\times10^{10}$\msun\ and the host is forming stars at a rate of $7\times10^{3}$ \msun/yr. The
quasar clearly sits in a region of
high overdensity, at the convergence of many filaments. The dense gas surrounding the quasar has a relatively high neutral fraction compared to the lower density material in the IGM. This is because we have not included the quasar local ionizing radiation (proximity effect) in this plot. The neutral fraction in Figure \ref{astridq} is the one computed as the simulation is run, using a uniform UVBG. In Section \ref{astridspec} we include the local quasar radiation in the \astrid\ 
\lya\ forest in post-processing.

\begin{figure}
    \centering
    \includegraphics[width=\linewidth]{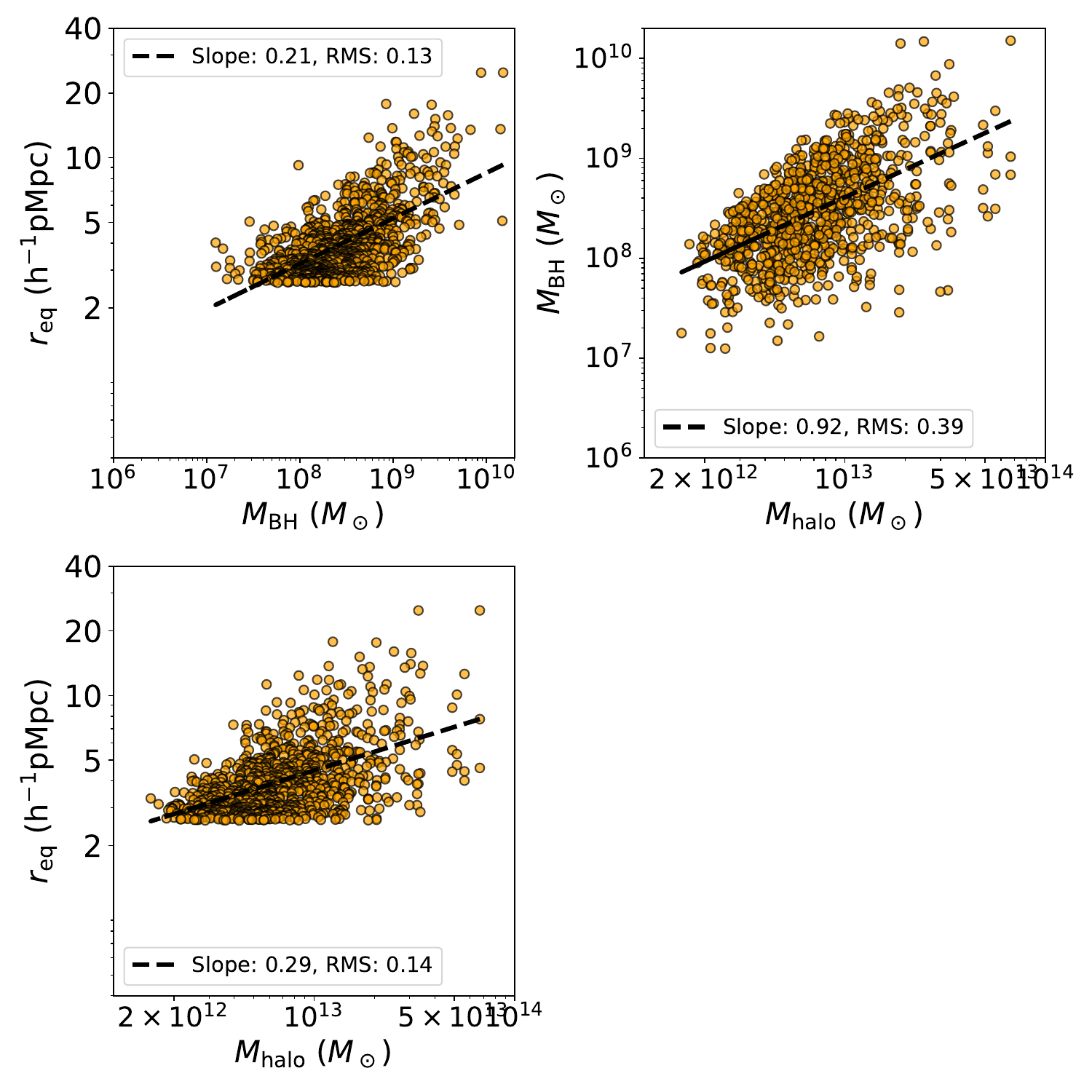}
    \caption{Properties of the black holes at redshift $z=3$. We show the top 1000 quasars by bolometric luminosity. In the top left we show radiation equality radius $r_{\rm eq}$ (Equation \ref{req_eqn})
 vs black hole mass,
    in the top right we show black hole mass vs halo mass, and the bottom $r_{\rm eq}$ vs halo mass.}
    \label{bhproperties_lbol}
\end{figure}

\begin{figure}
    \centering
    \includegraphics[width=\linewidth]{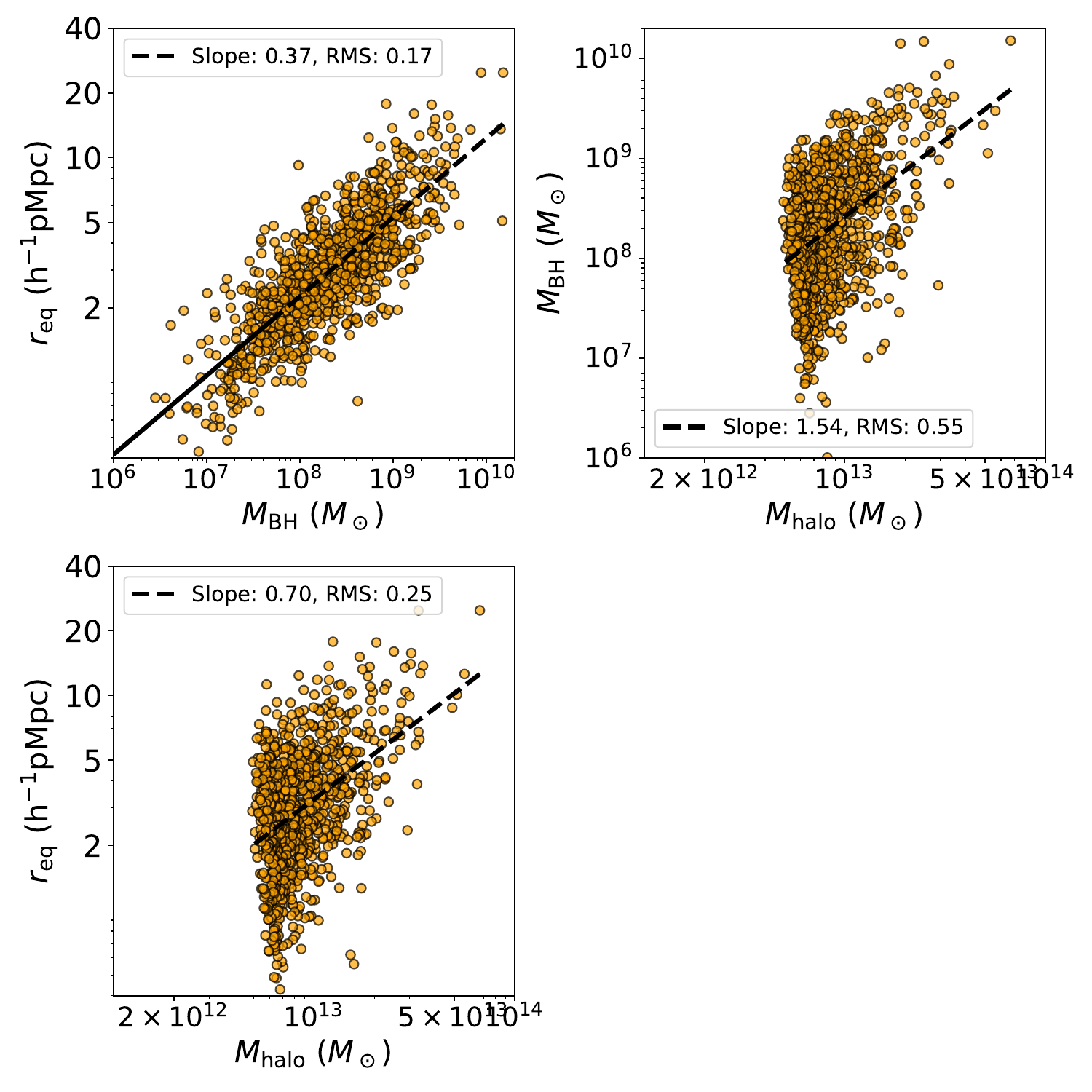}
    \caption{Properties of the black holes at redshift $z=3$. We show the top 1000 halos by mass.
    In the top left we show radiation equality radius $r_{\rm eq}$ (Equation \ref{req_eqn})
 vs black hole mass,
    in the top right we show black hole mass vs halo mass, and the bottom $r_{\rm eq}$ vs halo mass.
    }
    \label{bhproperties_mhalo}
\end{figure}

The UV luminosity function of the quasar population in \astrid\ is shown in 
Figure 5 of \cite{ni22}, and there is reasonable agreement with observational data. Here we select quasars from \astrid\ for our tests of the halo model. Because
\astrid\ contains a model for black hole seeding and tracks their growth through accretion
of gas and mergers with other black holes, it is useful to examine the relationship
between quasars and their host halos. This includes the black hole accretion rate (and therefore quasar luminosity), which early influential 
analyses (e.g., \citealt{martini01}) assumed were tightly coupled for ease of interpretation. Because of this we select two populations of quasars from \astrid, the first being the brightest 
1000 in the simulation and the second being those hosted by the most massive 1000 halos.

In Figure \ref{bhproperties_lbol} we show some properties of the brightest 1000 quasars in \astrid\ by 
bolometric luminosity. We have converted the quasar bolometric luminosity  to $r_{\rm eq}$ values by first computing the 
rate of emission of
hydrogen ionizing photons (\(\dot{N}\)) following 
 \cite{chen2021}: We adopt a power-law spectral energy distribution (SED) spanning 1450\,\AA\ to 912\,\AA, with \(L_{\nu} \propto \nu^{-\alpha}\) and \(\alpha = 1.5\), where the normalization is determined by \(L_{\mathrm{bol}}\). Applying the bolometric correction from \citet{fontanot12} then yields \(M_{\mathrm{UV}}\):
\begin{equation}
    M_{\mathrm{UV}} = -2.5 \log_{10} \frac{L_{\mathrm{bol}}}{f_{\mathrm{B}}\mu_{\mathrm{B}}}
    + \Delta_{\mathrm{B,\,UV}} + 34.1,
\end{equation}
with \(f_{\mathrm{B}} = 10.2\), \(\mu_{\mathrm{B}} = 6.7 \times 10^{14}\,\mathrm{Hz}\), and \(\Delta_{\mathrm{B,\,UV}} = -0.48\).
Following observations that suggest large escape fractions (\citealt{Eiler2021_young_qso}; \citealt{stevans_2014_qso_fesc}; see also \citealt{cristiani16}, who measure a mean escape fraction of 75\%-82\% at $z=3.8$ depending on the assumed continuum shape), we set the quasar’s escape fraction to \(f_{\mathrm{esc}} = 100\%\). The total ionizing photon rate is then obtained by integrating over energies above 13.6\,eV,
\begin{equation}
    \dot{N} = \int_{13.6\,\mathrm{eV}}^{\infty}\frac{L_{\nu}}{h\nu}\,d\nu,
\end{equation}

which is equivalent to $\log_{10}(\dot{N})=-0.4 M_{\rm UV}+46.366$ in our case.
We then use Equation \ref{req_eqn} to compute the radius $r_{\rm eq}$, assuming a hydrogen photoionization
rate from the UVBG, $\Gamma_{\rm HI}=10^{-12}$s$^{-1}$. This value of  $\Gamma_{\rm HI}$ is consistent with the results
of e.g., \cite{dall08}, \cite{kulkarni19},   and also the value used at $z=3$ when running the \astrid\ simulation
itself (taken from \citealt{fg20}).

We plot $r_{\rm eq}$ (in proper h$^{-1}$Mpc) against black hole mass and halo mass, and as
expected there is a sharp cutoff for low values of $r_{\rm eq}$. There is a relationship between
the quantities with a significant scatter. A power law fit has a slope of $n=0.29$. The values of $r_{\rm eq}$ for the faintest of the 1000 quasars are $r_{\rm eq}=3\phmpc$. This is relatively small compared
to the $\sim 10 \phmpc$ seen in samples of quasars used to measure the proximity effect
(e.g., \citealt{dall08}). This is because even though \astrid\ has a volume of $(0.25 h^{-1} \rm{cGpc})^{3}$ the mean space density of quasars with $M_{\rm UV} < -25$ at $z=3$ is $\sim 10^{3}$ per \hcgpccc (both in the simulation \citealt{ni22} and observationally, \citealt{kulkarni19}). This represents only $31$ in the \astrid\ volume so the majority of the 1000 represented in  Figure \ref{bhproperties_lbol} are comparatively faint. Returning to Figure
\ref{bhproperties_lbol}, we can see that there is a correlation between black hole mass and halo mass, with a scatter of about 0.4 dex. We note that  black hole seeding prescriptions can have strong effects on the black hole - halo mass mass relation (see e.g., \citealt{kho25}). 

In Figure \ref{bhproperties_mhalo} we show the population of quasars chosen by taking the most massive 1000 host halos in \astrid. We can see the correlations between properties, but these have a
different appearance, due to the cutoff in $r_{eq}$ not being present in this sample. Although the populations appear to be somewhat different for these
two criteria, we find in Section \ref{qrhof} below that the halo model parameters
for them are very similar and our analysis is insensitive to which population is
used.

\subsubsection{The quasar-mass density cross-correlation function in \astrid}
\label{qrhof}

Because our halo model for the proximity effect starts from the matter density profile around quasars (Equation \ref{eqnfw}), it is useful to see how well it can describe the density around the \astrid\ quasars. This will also give us an idea of the magnitude of the various halo model parameters. 

\begin{figure}
    \centering
    \includegraphics[width=\linewidth]{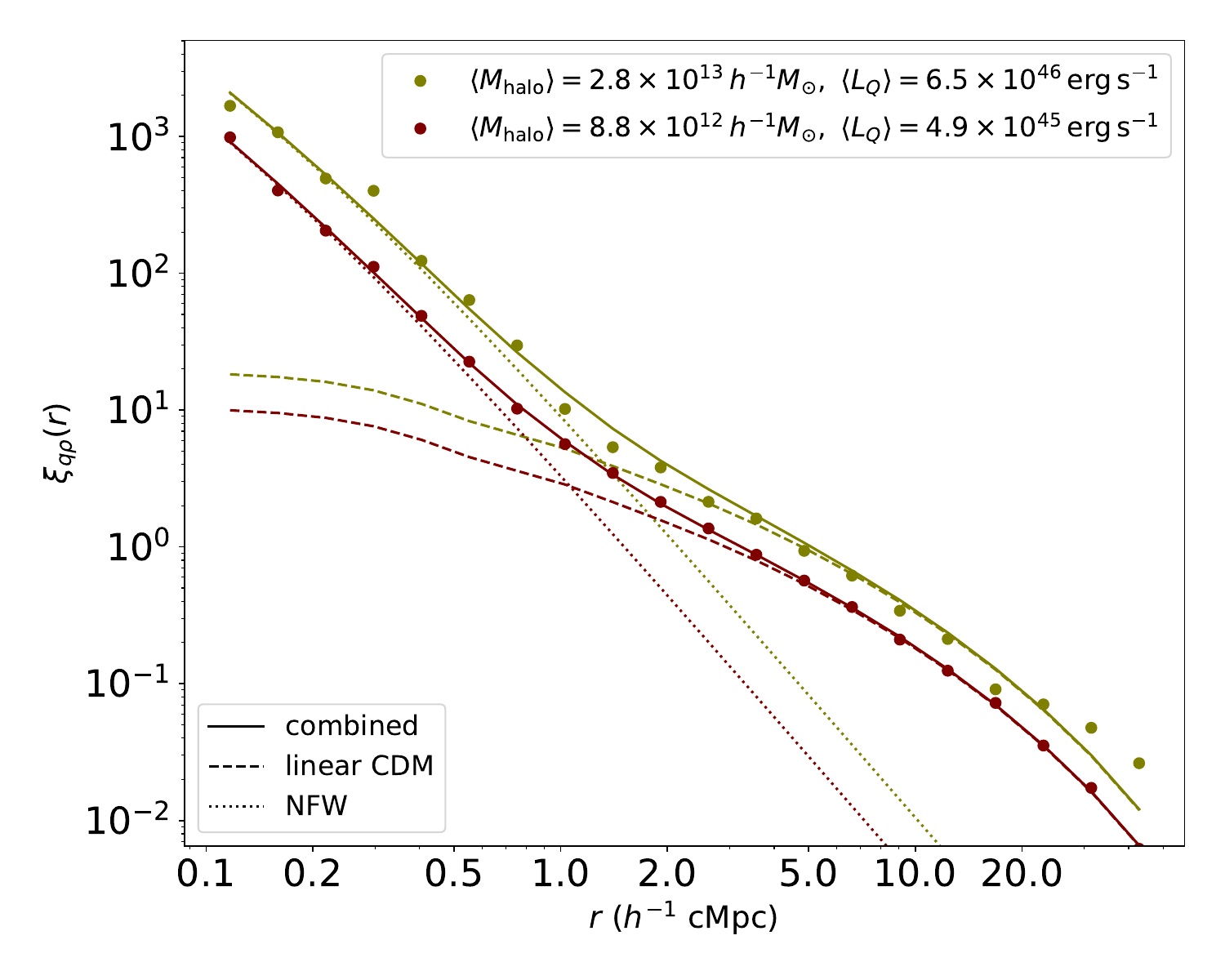}
    \caption{The quasar-mass cross-correlation function in \astrid\ together with halo model fits. The simulation data are shown as points, and the NFW and linear CDM fits are shown separately as dotted and dashed lines. The combined fits (NFW+linear) are shown as solid lines. We show results for two sets of quasars (the top 10 most luminous and the top 1000 most luminous) together with their mean halo mass and mean bolometric luminosity, as stated in the Figure legend}
    \label{xiqrho}
\end{figure}

We have sampled the gas particles in the simulation at z=3, randomly picking 0.001 of the total.  In Figure \ref{xiqrho} we show the quasar-gas density cross-correlation function, $\xi_{q\rho}$ as a function of scale, for two samples of quasars in \astrid. These samples are the most luminous 1000 objects and the most luminous 10 in the volume  The mean bolometric luminosity of the brighter sample is 13.3 times that of the other set of quasars, and the average host halo mass is a factor of  3.2 times more. The simulation datapoints
trace out a similar shape, with the brighter quasars
having a higher amplitude of clustering. The transition
between the one-halo and two-halo parts of the curve are readily visible from the change in slope on scales
$r \sim 1 \hmpc$.

We carry out a fit of the
halo model curve of Equation \ref{eqnfw} to the simulation points. 
 The fit is unweighted least‐squares in log space using the Python function \texttt{curve\_fit}, and assumes identical errors for all data points
(for our actual proximity effect fits later in the paper we will compute and
use the full covariance matrix). The best fit lines are shown in Figure \ref{xiqrho}, as well as the individual
components, the NFW term and the linear CDM term. The brighter sample (top 10 most luminous quasars) has a best fit linear bias of $b_{q}=4.0$.

\begin{figure}
    \centering
    \includegraphics[width=\linewidth]{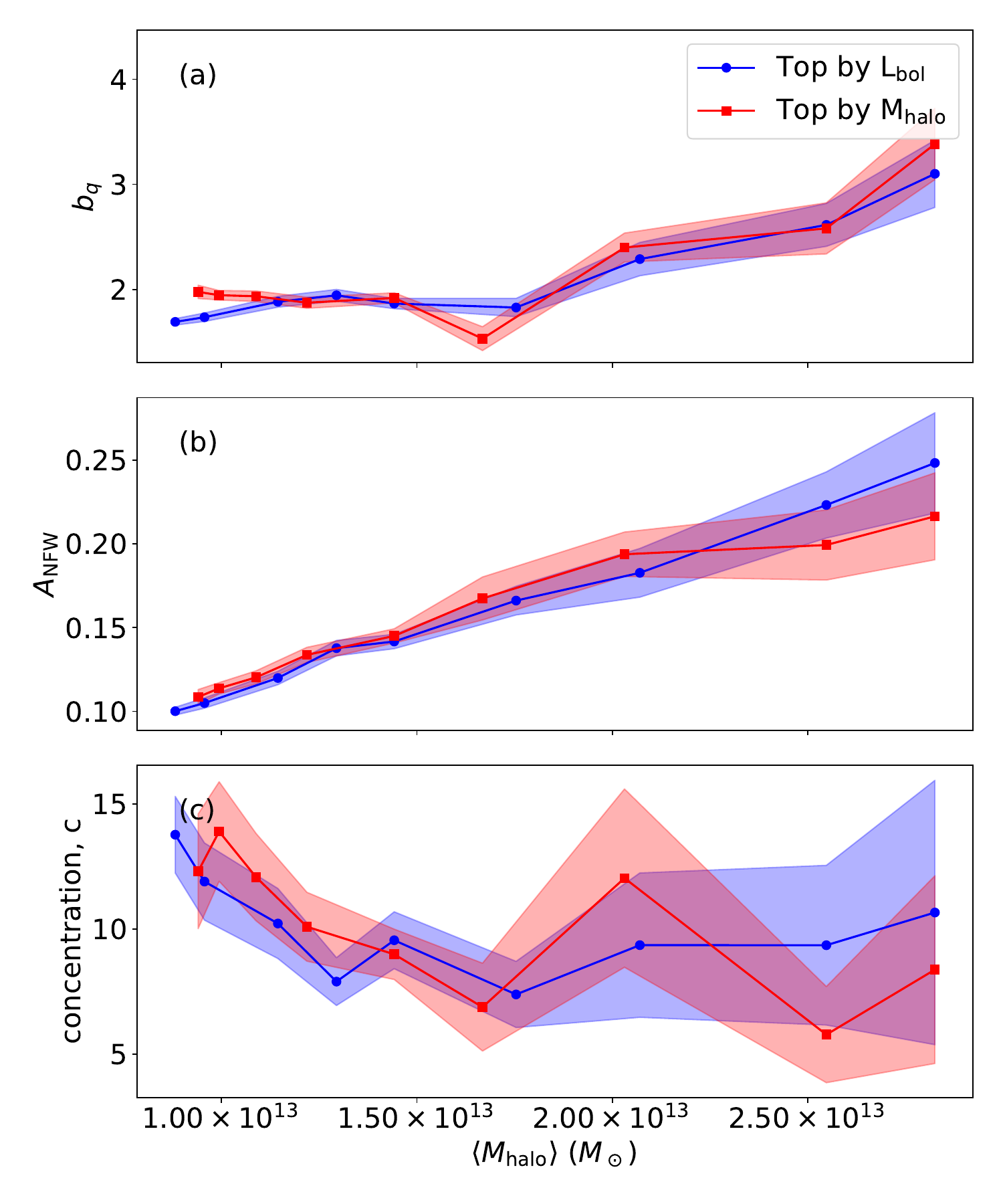}
    \caption{Halo model parameters (see Equation \ref{eqnfw})
    fit to the quasar-mass cross-correlation function in \astrid. We show the halo parameters as a function of halo mass. The results are shown for two different samples, the top 1000 quasars by bolometric luminosity (in blue) and top 100 by host halo mass (red). The shaded areas show the $\pm 1 \sigma$ 
    uncertainty in the fit parameters.
    } 
    \label{qrhoparams}
\end{figure}

Both the NFW and linear CDM terms have a higher
amplitude for the brighter sample, and we quantify this further
by carrying out fits as a function of halo mass for
the top 1000 luminous quasars and the top 1000 massive
host halo quasars. In this analysis we bin the objects into 9 bins of halo mass and compute the fits for objects in each bin.
The fits use the standard Levenberg–Marquardt nonlinear least‐squares routine from \texttt{curve\_fit}, treating each point as having equal weight. The resulting parameter covariance matrix used to compute the one-$\sigma$ confidence regions is the approximate inverse of the Hessian at the best‐fit point, as returned by the fitting algorithm.
The results are shown in Figure \ref{qrhoparams}, where the parameters $b_{q}$, $A_{\rm NFW}$
and $c$ are shown as a function of mass. 
Both sets of quasars (luminous and massive halos) behave very similarly, with error bars on all parameters overlapping for nearly the entire range. The plots are on a linear scale, and we can see that $b_{q}$  is flat or rises slowly at first  from $b_{q}\sim 2$ for lower mass halos and then faster. This is as expected from 
other calculations in the context of the halo model (e.g., \citealt{seljak04}). The amplitude of the NFW portion rises more linearly with halo mass, and the concentration $c$ falls at first and the remains roughly constant at $c\sim 5-10 $ for most of the range.

\subsubsection{Quasar luminosities compared to observational data}
\label{quasarlumobs}
As mentioned previously, the volume of \astrid\ is such that it contains a few quasars, but is not large enough to cover a bright sample of the quasars selected by a 
survey such as DESI (\citealt{desi16,martini25}). DESI covers a sky area of $13000$ square degrees, and so the volume between redshifts $z=2.5$ and $z=3.5$ is 56 \hcgpccc. This is approximately 3600 times volume of \astrid. This means that if we would like to model more than a handful of the bright quasars in DESI using \astrid\ objects, we will need to make an approximation,
described below.

Using the $z=3$ quasar bolometric luminosity function \cite{shen20}, we compute that the median $L_{\rm bol}$ for the brightest 1000 quasars in DESI between $z=2.5-3.5$ will be
$L_{\rm D}=3.5\times 10^{47}$ \ergs. This translates to an $r_{eq}$ value of 10 \phmpc\ (Equation \ref{req_eqn}). We also
compute the median $L_{\rm bol}$ for the brightest 1000 quasars in \astrid, finding
$L_{\rm A}=7.0\times 10^{45}$\ergs. In our analysis of the proximity effect in \astrid\ we therefore scale up the individual $L_{\rm bol}$ values of quasars in \astrid\ by a 
uniform factor of $L_{\rm D}/L_{\rm A}$. The median
value of $r_{\rm eq}$ for the 1000 quasars in our scaled-up \astrid\ sample is then
$r_{\rm eq}=10\phmpc$. When carrying out our fits of the proximity effect using this scaled-up sample we will now be testing over a range relevant to observational data from a large quasar survey such as DESI. Unfortunately it isn't easily possible to adjust the physical
environment of the quasars so that they lie in rarer, denser environments (and so have 
higher $b_{q}$ values). However, as we shall see, the fits to $r_{\rm eq}$ and $b_{q}$ have minimal degeneracy, so that we can reasonably believe that the simulation comparisons are still a fair test of the halo model. We have also carried out a test (described fully later) where we assign much lower luminosities to the subset of quasars in the most massive halos in \astrid\ (equivalent to their actual expected luminosities in DESI) to check that the halo model proximity effect fits still give good results.

We use scaled quasar luminosities in Section \ref{anasim}. In Section \ref{largefield} where we map out the ionizing background inferred from the proximity effect on large scales, we use a different, larger (dark matter) simulation. In this case we use the actual luminosities associated with halos of a given mass in \astrid\ to make a properly representative radiation field.

\subsection{Simulated \lya\ forest spectra in \astrid}

\label{astridspec}

We make \lya\ forest spectra that are parallel to one of the axes of the simulation and start at the positions of the 1000 quasars in our \astrid\ samples. The spectra are generated following standard methods commonly used in SPH simulations (e.g., \citealt{hernquist96}). Briefly, the neutral hydrogen density is mapped onto spectral pixels based on the contributions from SPH kernels of intersecting particles. This approach is similarly applied to other physical properties, which are weighted according to the neutral hydrogen density,  directly correlated with optical depth for absorption. For each line of sight, the simulation spectra contain several key physical quantities relevant to the modeling: the optical depth of neutral hydrogen absorption in real space,the peculiar velocity component along the sightline, the temperature,
T, and the normalized baryon density in units of the cosmic mean density.

In order to include the effect of local quasar radiation from the quasar itself, we modify the
optical depth $\tau$ along the \lya\ forest sightline according to Equation \ref{tauprox}. Here for each spectrum we assume that the radiation comes from either the UV background radiation or
radiation from the quasar itself. We do not compute the contribution from other quasars in the volume, modeling that instead using the uniform background (this assumption is 
relaxed in Section \ref{largefield} where we look specifically at fluctuations in the UV background).
The final observable in the \astrid\ spectra, the transmitted flux $F=e^{-\tau}$ is obtained after 
convolving $\tau$ with peculiar velocities and thermal broadening. 

The spectra are computed using pixels of width 5 kms$^{-1}$, but we rebin
to 200 kms$^{1-}$ pixels before analyzing
them. For comparison, the DESI 0.8 \AA pixels are 50 \kms in width at redshift $z=3$  (\citealt{ramirez24}). Although
our fiducial analyses are with noiseless spectra,  in Section \ref{indivspec} we explore the impact of adding DESI-like noise to the
simulated spectra.

\section{Analyzing simulated spectra using the halo model}
\label{anasim}

\subsection{Mean spectra}
\label{meanspectra}

\begin{figure}
    \centering
    \includegraphics[width=\linewidth]{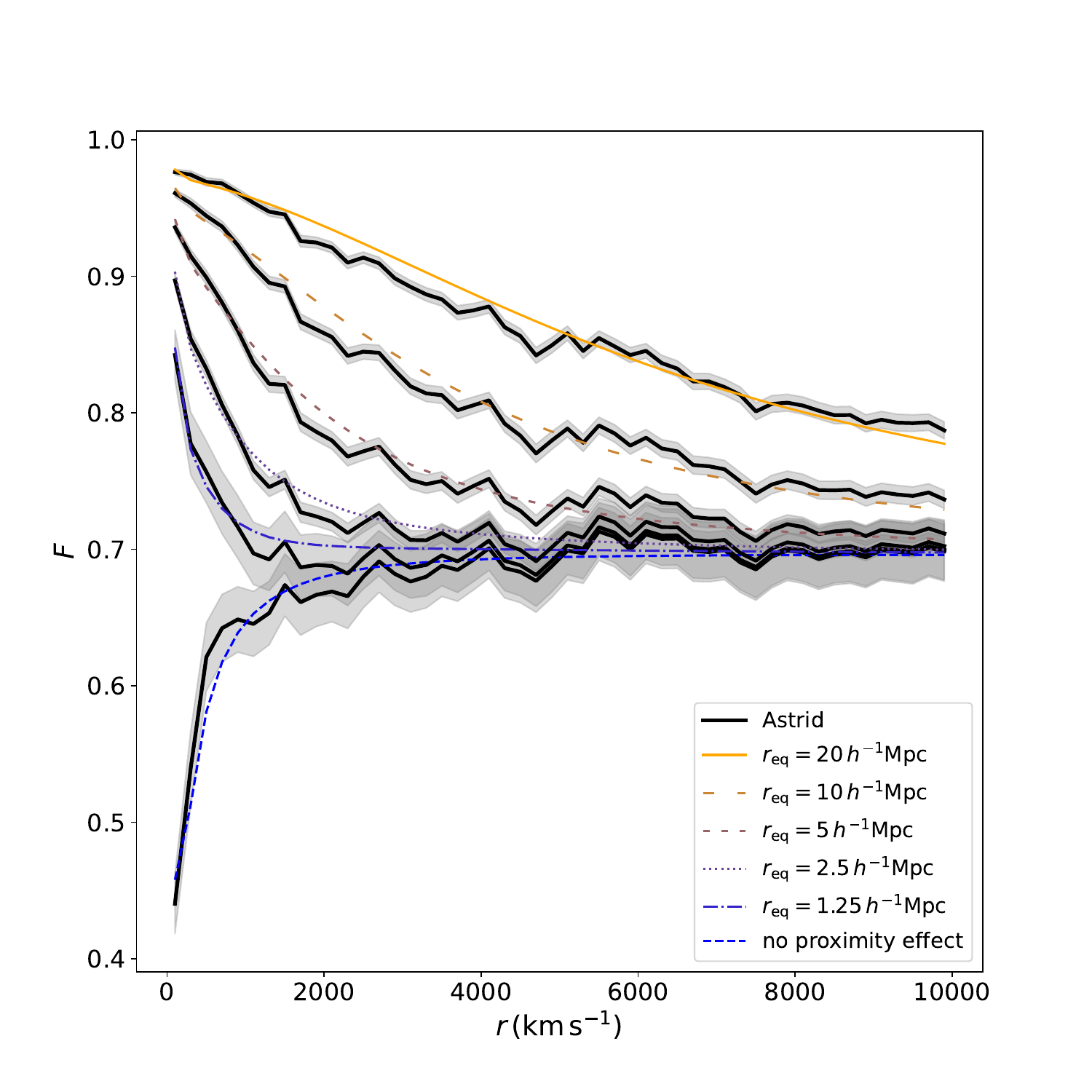}
    \caption{The effect of quasar radiation, parametrized by $r_{\rm eq}$. We show the halo model
    of \lya\ transmitted flux as a function of separation $r$ from the quasars. We also show results from the \astrid\ simulation for the top 1000 quasars by halo mass, where the
    $r_{\rm eq}$ values have been set to different values given in the legend. The simulation results are shown as solid lines, with a gray band representing the standard deviation
    computed from the 1000 spectra. The best fit halo models are shown as dotted and dashed lines. The results labelled "no proximity effect" are for the case where we have not
    included the local radiation from the quasar, only the UV background.}
    \label{halosimexample}
\end{figure}

Our observable is the profile of \lya\ forest transmitted flux, $F$ on the blue side of the
\lya\ emission line in quasar spectra.
When analyzing spectra from simulations or observations, we can look at individual 
spectra or we can average many spectra (perhaps in different bins of quasar luminosity).
We choose to do the latter first, and in Figure \ref{halosimexample} we show some
example $F$ profiles. In this case we have taken the entire sample of the top 1000 quasars by
host halo mass and given them the same luminosity (parametrized by the $r_{\rm eq}$ values 
given in the legend) before computing the mean $F$ profile as a function of distance $r$.
We have repeated this for a range of $r_{\rm eq}$ values, from $r_{\rm eq}=1.25$ \phmpc\ to
$r_{\rm eq}=20\phmpc$\ . We have also computed the $F$ profile for the case without the local
contribution to the ionizing radiation from the quasar itself. This is the bottom-most line,
and we can see that it is the only one where the transmitted flux monotonically decreases
as we approach $r=0 \kms$. The curve labelled $r_{\rm eq}=10\phmpc$ is approximately what we
expect the averaged profile for the top 1000 most luminous DESI quasars should look like.

\begin{figure}
    \centering
    \includegraphics[width=\linewidth]{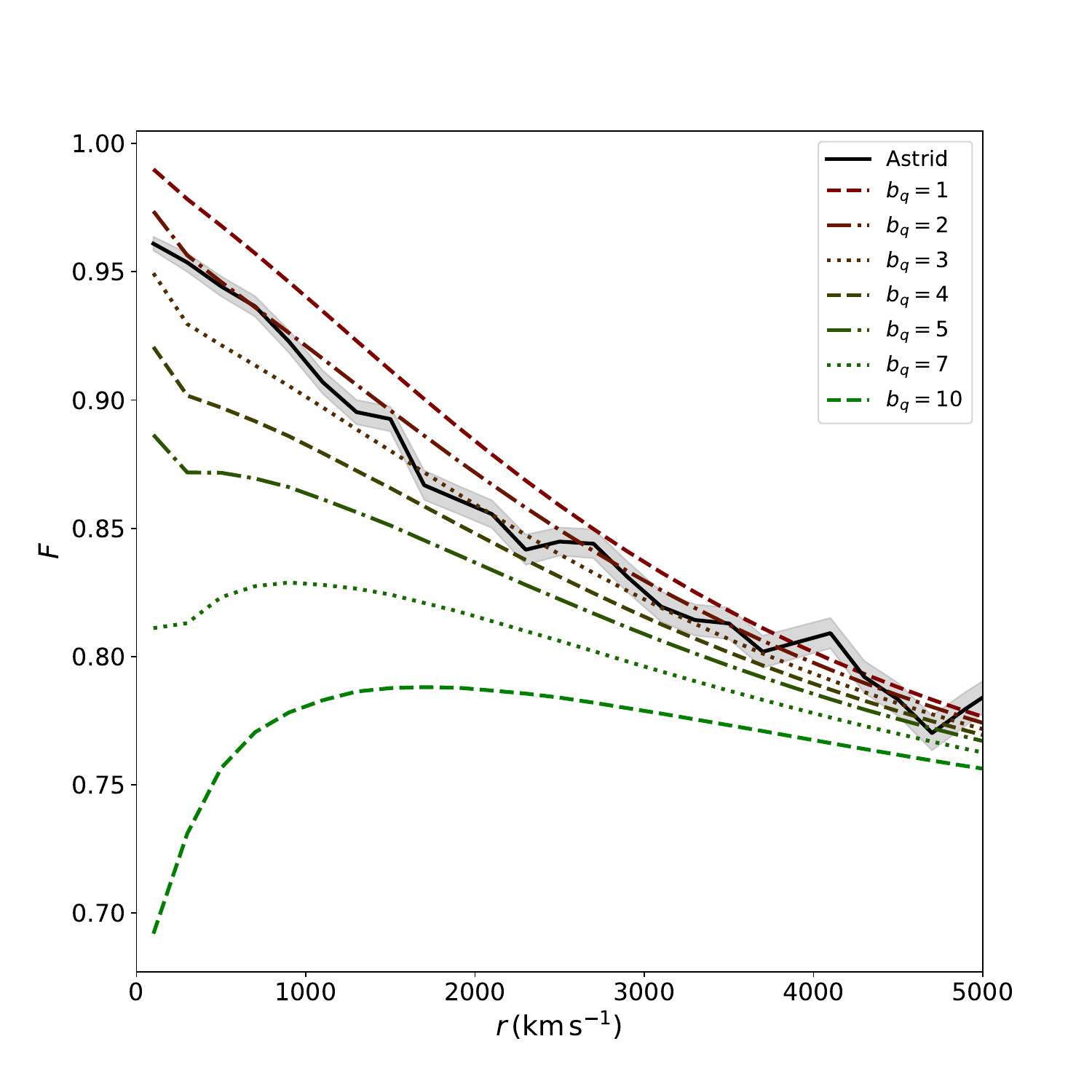}
    \caption{The effect of linear bias, parametrized by $b_{\rm q}$. We show the halo model
    of \lya\ transmitted flux as a function of separation from the quasars. We also show results from the 
    \astrid\ simulation for the top 1000 quasars by halo mass. For all models (and in the simulation) the value of
    the ionizing radiation parameter $r_{\rm eq}=10 \phmpc$ }
    \label{halosimexamplebq}
\end{figure}

We use the halo model to compute theoretical $F(r)$ profiles for the same $r_{\rm eq}$ parameters that were used in the simulation
spectra. We use a value of $b_{q}=2.0$, which is
the relevant median bias parameter for the  1000 most luminous quasars in \astrid\ as we have seen in the analysis in Section \ref{qrhof} (see Figure \ref{qrhoparams}).
In Figure \ref{halosimexample} we can see that these models describe the simulation data quite well, both in shape and
amplitude.

In Figure \ref{halosimexamplebq}, we show the halo model $F(r)$ curves for different values of the $b_{q}$ parameters (all
for $r_{\rm eq}=10 \phmpc$. As we would expect, the higher values
of $b_q$ diminish the proxmity effect, so that the monotonic
rise of $F(r)$ is halted for the most extreme values ($b_{q}=7$ and higher).
The simulation curve for the top 1000 halos by halo
mass is also shown on the plot. We will now see how well the
differentiation between curves for different values of
$b_{\rm q}$ and $r_{\rm eq}$ translate into fits for those
parameters.

\subsubsection{Model fits}

\label{modelfits}

We compute the covariance matrix $C(r_{i},r_{j})$ of the $F(r)$ bins by computing
$F(r)$ for each of the $N=1000$ sightlines in \astrid:

\begin{equation}
C(r_i, r_j) = \frac{1}{N - 1} \sum_{n = 1}^{N}
\Bigl[F_n(r_i) - \overline{F}(r_i)\Bigr]
\times \Bigl[F_n(r_j) - \overline{F}(r_j)\Bigr],
\label{covmatrix}
\end{equation}

where

\begin{equation}
\label{eq:mean}
\overline{F}(r_i) = \frac{1}{N} \sum_{n=1}^{N} F_n(r_i).
\end{equation}

\begin{figure}
    \centering
    \includegraphics[width=\linewidth]{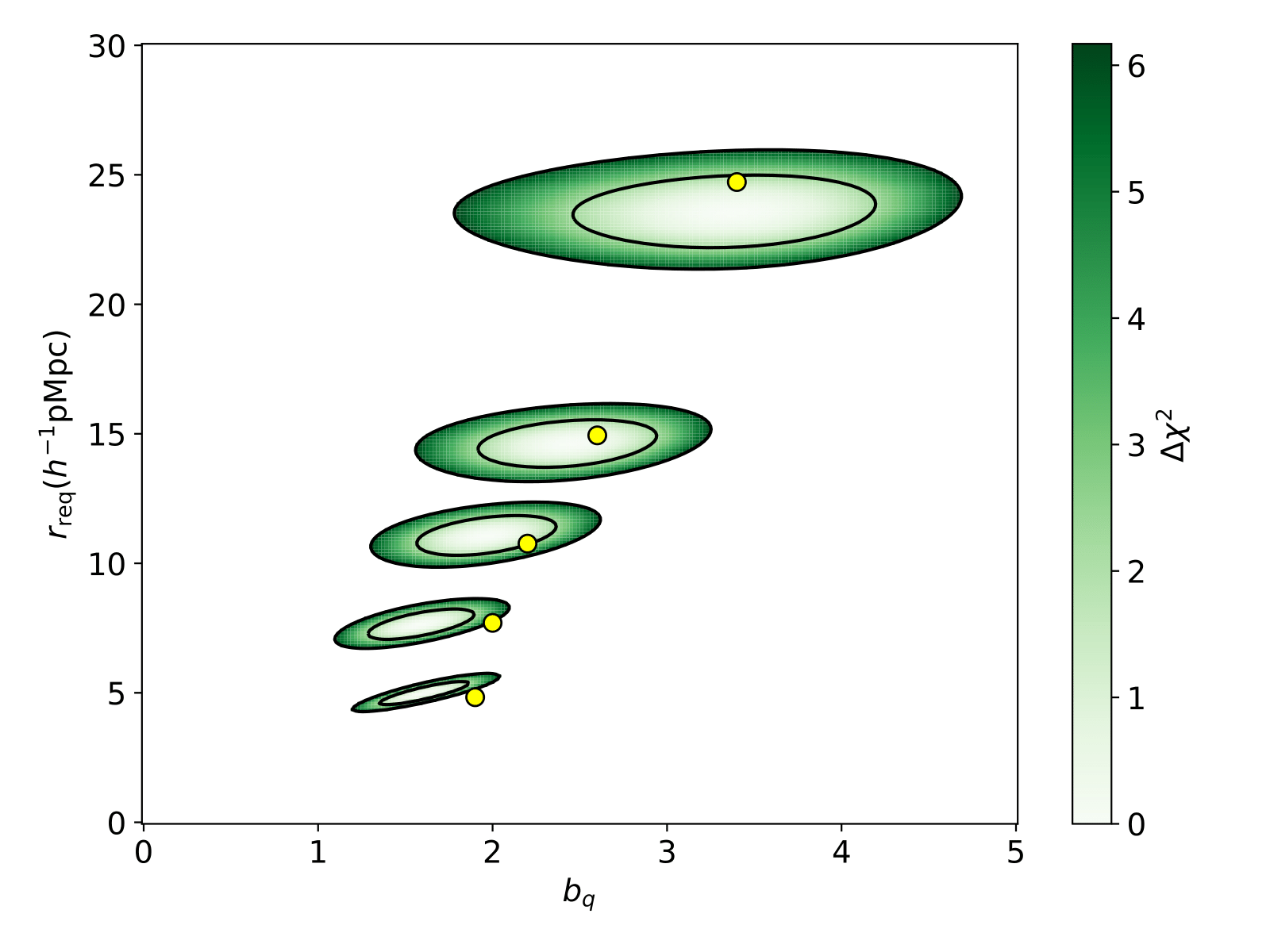}
    \caption{Confidence contours (1 and 2 $\sigma$ of model parameters, $b_{q}$ and
    $r_{\rm eq}$ fit to the \lya\ forest proximity effect from sets of 200 quasars (ranked by halo mass) in \astrid. The top contours show results for the most massive 200 host halos, the next contours down the next 200 and so on. The values of the $r_{\rm eq}$ parameter used in the simulation and the $b_{q}$ measured from the quasar-mass cross-correlation (the "true values" of the parameters) are shown with yellow dots.
    }
    \label{multiconts}
\end{figure}

We use the halo model to compute theoretical $F(r)$ profiles and
perform a chi-squared fit of the grid of models to simulation data. We order the simulation spectra by host halo mass, and then split the sample of 1000 spectra into 5 sets of 200 halos. We multiply the covariance matrix from Equation \ref{covmatrix} by 5 to
account for the smaller sample.
We compute the mean $F(r)$ for each set, and then 
 carry out the chi-squared fit. The 1 and 2 $\sigma$ confidence
 contours on the fit parameters are plotted in Figure \ref{multiconts}. We also show the actual values of $r_{\rm eq}$
 and $b_{\rm q}$. We can see that fits are reasonable, lying within the 1 and 2 $\sigma$ contours in most of the samples.

 It is interesting that even though increasing 
 $b_{\rm q}$ and decreasing $r_{\rm eq}$ have similar effects
 (Figures \ref{halosimexample} and \ref{halosimexamplebq}) there is not much covariance
 between the parameters in the contours in Figure \ref{multiconts}. This means that it will be possible to make measurements of $r_{\rm eq}$ and constrain the proximity
 effect even if the constraints on $b_{\rm q}$ are  weaker. 

  \begin{figure*}
    \centering
    \includegraphics[width=0.9\linewidth]{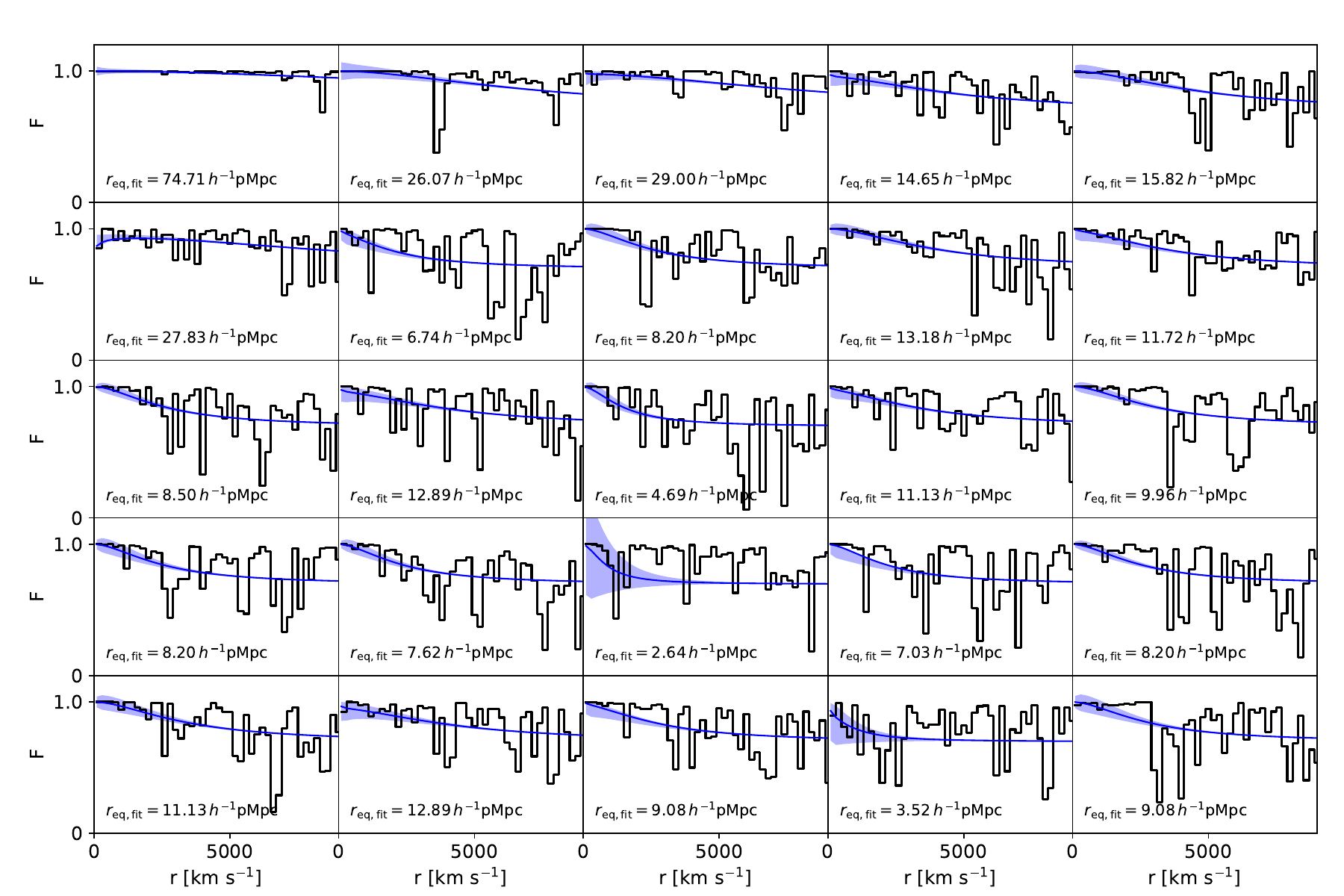}
    \caption{
Halo model fits to individual quasar spectra. We show the transmitted flux, $F$ for 25 of the most luminous 1000 \astrid\ quasars (every 40th spectrum is plotted, in order of quasar luminosity). The quasars are positioned on the left of each panel, at $r=0$ kms$^{-1}$ and the \lya\ forest spectra are shown as
black lines. The best fit halo model curve is shown as a blue
line, with the plus and minus one sigma best fit curves shown as shaded regions. The best fit value of $r_{\rm eq}$ is
given in each case.
}
    \label{gridofplots}
\end{figure*}

On the other hand, if we compute the $1 \sigma$ error on $b_{q}$ from the 
confidence contours, we find it to be $0.50$ averaged over the two samples of 200 quasars with $b_{q}\sim 2.5$ and $b_{q} \sim 3.5$, or a fractional uncertainty of 0.17. We can  gauge how this converts into a constraint on the mean host halo mass by making use of the halo mass-$b_{q}$ relation plotted in Figure \ref{qrhoparams}. In that plot we see that the slope of 
$\left< M_{\rm halo} \right>$ vs $b_{\rm q}$,
$
\left(\frac{\Delta b_{\rm q}}{b_{\rm q}}\right)
\left(\frac{\Delta M_{\rm halo}}{M_{\rm halo}}\right)^{-1}
= 2.0$
 for $b_{q}=3$. This translates
the fractional uncertainty of 0.17 in $b_{q}$ from 200 quasars into a
fractional uncertainty of 0.09 in the host halo mass. We can compare this to Figure 14 of \cite{faucher08}, where the proximity effect was shown to recover the host halo mass from a sample of 100 quasars at $z=3$ with a fractional error of 0.19. Of course, as in the initial analysis of \cite{faucher08} we assume perfect data 
with no noise and perfectly fitted continuua. We discuss this further in  Section  \ref{discussion}.

Because of the small volume of \astrid\ compared
to DESI, we have had to boost the luminosities of the quasars by a factor of 50 (Section \ref{quasarlumobs}). The expected relationship between halo mass and $r_{\rm eq}$ in DESI will therefore be different from 
that used in our tests. Ideally we would have a much larger
simulation that includes all DESI quasars, but this is not currently
feasible. We can however test our proximity effect fits in a very different regime by remaking spectra for the  \astrid\ quasars in 
the most massive halos, but without the factor of 50 boost. We have 
done this with the top 20\% of quasars by halo mass in Astrid. The
input $r_{\rm eq}$ value in the simulation in this case is 3.5 \phmpc\ (compared to 25 \phmpc\ used for the top 20\% sample in Figure \ref{multiconts} ).  We fit the halo model proximity effect to the mean flux-distance profile, and find $r_{\rm eq}=4.2 \pm 0.6$ \phmpc, which is in 
reasonable agreement. The expected $b_{q}$ for these halos (from Figure \ref{multiconts}) is $b_{q}=3.5$, and our fit returns
$b_{q}=2.8 \pm 0.8$. This shows that
no major issues have been discovered so far, although of course a full test will have to wait for yet more massive halos in a much larger future simulation.

\subsection{Individual spectra}
\label{indivspec}
The halo model fits of the proximity effect can also be carried out on individual
spectra. As \cite{dall08} have shown, the effect is noticeable even in this case, and so we  perform the same fits as in Section \ref{modelfits} except that we multiply the covariance
matrix $C_{ij}$ by 1000 as it was computed using that many sightlines. In Figure \ref{gridofplots} we show results for 25 of these spectra. They are ordered in terms
of luminosity, and  span the 1000 quasars, starting with the most luminous and then skipping through every 40  to end on the least luminous. Comparing the fits to the spectra it is obvious that the fluctuations due to the density inhomogeneities
in \lya\ forest gas are large and that the individual profiles are nowhere
as smooth as the averaged simulation profiles in Figure \ref{halosimexample}. To the extent that this is encoded in the covariance matrix, this shouldn't be a problem, and has the advantage that instrumental or observational noise is unlikely
to be an overwhelming  factor in the fits. For example, DESI spectra have a mean S/N
of order unity in the \lya\ forest
(\citealt{desiedr}), so that the noise in fitting would be not very different from that due to the
real structure that we are simulating. Later we carry out a simple test where we add some noise to the spectra to approximate that in DESI observations, showing that this is indeed the case.

The fits themselves are shown as smooth lines in Figure \ref{gridofplots}, and the
best fit value of $r_{\rm eq}$ is also shown. Among the quasars plotted these range
from an extreme value of $r_{\rm eq}=74.71 \phmpc$ for the brighest quasar to 
$r_{\rm eq}=3.52$\phmpc for one of the faintest. We also show as a shaded region the curves which correspond to the best fit  $r_{\rm eq}$ values plus or minus 
1 $\sigma$ confidence, computed from the $\Delta \chi^{2}$ differences.

\begin{figure}
    \centering
    \includegraphics[width=\linewidth]{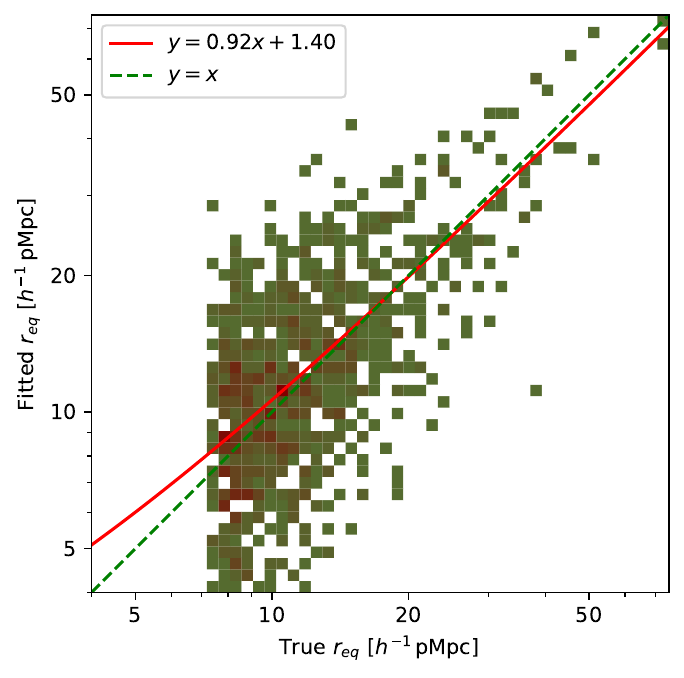}
    \caption{Results of fits to individual quasar spectra. We show
    the best fit $r_{\rm eq}$ inferred from the halo model fit to the 
    proximity effect in \astrid\ against the true value for that quasar.
    (Results are for the brightest 1000 quasars in \astrid\ with luminosities scaled so that median $r_{\rm eq}$ is 10 \phmpc ). The red line shows the best fit straight line to the points and the green dashed line shows $y=x$.
    }
    \label{reqindiv}
\end{figure}

\begin{figure}
    \centering
    \includegraphics[width=\linewidth]{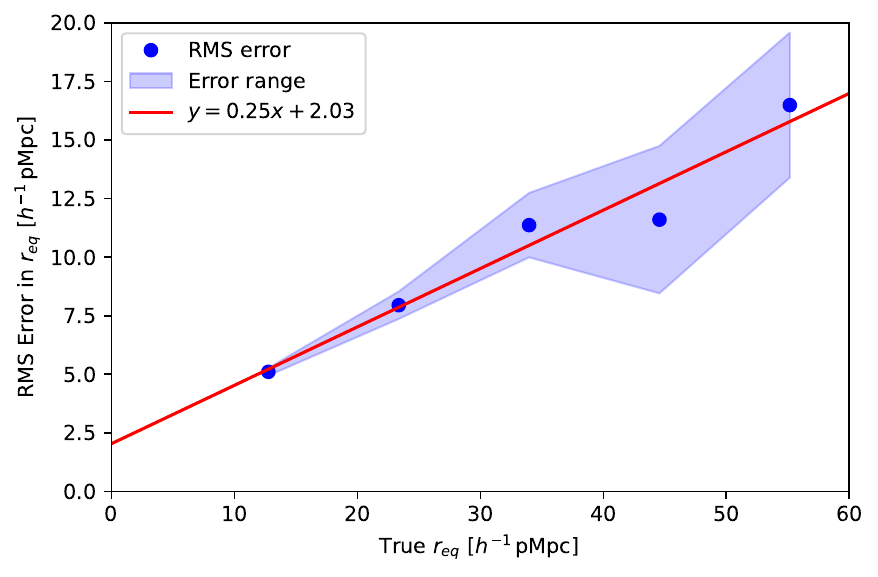}
    \caption{ RMS error in $r_{\rm eq}$ 
    for fits to individual quasars as a function
    of the true $r_{\rm eq}$ values. We show
    the bootstrap error on the error as a shaded
    region and the best fit straight line as a
    red line. 
    }
    \label{erroronreq}
\end{figure}

In order to check the recovery of $r_{\rm eq}$, in Figure \ref{reqindiv} we show the true values of $r_{\rm eq}$ and those from the proximity effect
fit. We carry out a chi-squared fit to the points,  finding that the fit is within $1 \sigma$ of the expected
$y=x$ line. Because it will be useful later to use the expected error on $r_{\rm eq}$ in another analysis (Section \ref{largefield}) we quantify the rms error
on $r_{\rm eq}$ from the fits in Figure \ref{erroronreq}. We can see that the
rms error vs $r_{\rm eq}$ is a straight line showing that the fractional error
is a constant 25\% plus a 2 \phmpc\ offset). Over the relevant range
this translates to a fractional error from $\sim 25\%$ for the brightest quasars,
to $\sim 50\%$ for the faintest.

At  this point, it is useful to gauge the likely effect of observational noise on the $r_{\rm eq}$ measurements. We have carried out a simple test, where we  approximately model the noise in DESI (based on the DESI Early Data Release, \citealt{ramirez24})
and add it to the simulated spectra. We follow the procedure outlined in \cite{shaw25}, estimating conservatively that the noise level in the continuum has a standard deviation
$\sigma=0.3$ in flux units 
for
DESI pixels of 0.8 \AA\ (50 \kms) in size. We add this
random noise as a Normally distributed variable to each pixel (where the pixels have been rebinned to this pixel size). We then carry out the fits of the halo model to measure $r_{\rm eq}$ values from each spectrum. Carrying out the same analysis that was used in the noiseless case (Figure \ref{erroronreq}) we find that the fractional error on $r_{\rm eq}$ is $0.23 r_{\rm eq}$ + 5.1 \phmpc. This translates to a
fractional error of $\sim 30\%$
for the brightest quasars and $\sim 65\%$ for the faintest. This is not very different from the noiseless case. Because we would like to avoid limiting our predictions to DESI (and because the noise model is crude and may also be overly conservative for
the brightest quasars), we will use the noiseless results for the rest of our analysis.

\begin{figure}
    \centering
    \includegraphics[width=\linewidth]{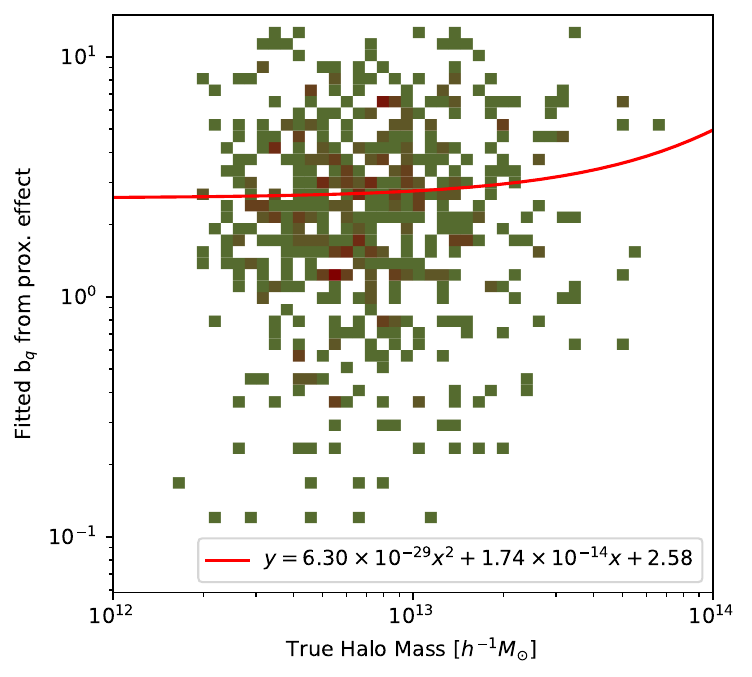}
    \caption{Results of fits to individual quasar spectra. We show
    the best fit $b_{q}$ inferred from the halo model fit to the 
    proximity effect in \astrid\ against the true host halo mass for that quasar.
    (Results are for the brightest 1000 quasars in \astrid\ with luminosities scaled so that median $r_{\rm eq}$ is 10 proper mpc/h ). The red line shows the best fit second order polynomial to the points.}
    \label{bqmhaloindiv}
\end{figure}

While we are most interested in the determination of $r_{\rm eq}$ for each quasar
(in order to constrain the UVBG intensity, as in e.g., \citealt{bajtlik88,scott2000,dall08}),
the halo model fit can in principle constrain the $b_{\rm q}$ parameter on an object by object basis, and hence the quasar host halo mass (\citealt{faucher08,kim04}). As we have seen from the elongated contours in Figure \ref{multiconts}, the constraints on $b_{q}$ are weaker than those on  $r_{\rm eq}$. Because it is not simple to determine the true $b_{\rm q}$ in the simulation for individual objects it makes most sense to do this with samples of quasars (as we have done in Section \ref{meanspectra}). Nevertheless,
to show the scatter involved, we plot in Figure \ref{bqmhaloindiv} the fitted $b_{\rm q}$ from the proximity effect against the true halo mass. We also show the best fit of a quadratic function to the points. From the results in Section \ref{meanspectra} (a 9\% error in halo mass from 
200 quasars), we expect a fractional error per quasar of about $\sqrt(200) \times 9\%= 130\%$. It will therefore only be practicable to use this method of constraining the halo mass by averaging over large samples of quasars.

\section{The large-scale radiation field from proximity effect measurements}

\label{largefield}

We have seen that the halo model can constrain $r_{eq}$ reasonably
well for individual quasars. The main use of the proximity effect
in the literature (e.g., \citealt{dall08}) has been to determine the 
ionizing background intensity, or equivalently the hydrogen photoionization rate $\Gamma$ given the quasar apparent luminosity. 
This can also be done using the halo model measurements  (by inverting Equation \ref{req_eqn} to find $\Gamma$). Rather than simply
measuring the mean value of $\Gamma$ as a function of redshift, we
could take advantage of the high space density and number of quasars in the surveys such as eBOSS (\citealt{dawson16}) and DESI (\citealt{martini25}) to measure spatial variations in the value of $\Gamma$. In this section we will test how well this can be done using a large
dark matter simulation. We use the results from \astrid\ to map quasar luminosities onto halo masses, as well as to determine the measurement errors in  $r_{\rm eq}$  from \lya\ forest proximity effect fitting.

\subsection{Simulating the large scale ionizing background radiation field}

The intergalactic hydrogen-ionizing radiation field at redshifts
$z=2-4$ is believed to be dominated by emission from quasars (see e.g., works from \citealt{haardt96} to \citealt{madau24}). The quasar radiation travels through space, gradually redshifting but
also being eventually absorbed by neutral hydrogen. The highest 
overall cross-section for this absorption comes from Lyman-limit
systems (\citealt{sargent89}), and the mean free path  of ionizing photons
is related to the number density and size of these systems. 
Both pure theoretical modeling and  more observationally derived estimates exist in the literature for this mean free path, or attenuation length, $r_{\rm att}$
For example.  \cite{theuns24} find an attentuation length
at $z=3$ of 125 proper Mpc,
(which is 88 \phmpc\ or 350 \chmpc, their Figure 5). \cite{daloisio18} estimates a similar value and the
observations of \cite{fumagalli13} point to  $100\pm29$ proper Mpc.

 At any point in space,
only a few bright quasar sources will be visible inside
the attenuation radius, and this will mean that the 
radiation field will fluctuate on the relatively large (tens of megaparsec) scales related to the mean
separation of sources.
These  fluctuations in the hydrogen-ionizing background
are expected to lead to a modulation of the 
clustering of the \lya\ forest. This has been studied
theoretically at least since \cite{zuo92} (see also 
\citealt{croft97b, croft04, mcquinn11}). Such fluctuations can affect clustering of the \lya\ forest on BAO scales (\citealt{gontcho14}) and are their potential effects are marginalized over
in BAO analyses alongside continuum fitting uncertainties. The large-scale radiation field fluctuations have yet to be detected directly, however, and the proximity effect offers a
possible route that we shall investigate.

No
cosmological hydrodynamic simulation of quasars 
yet exists which is large-enough to cover the Gigaparsec volumes we need to simulate the large-scale radiation field. We therefore turn to a combination of
a large dark matter simulation (\uchuu, \citealt{ishiyama21}), to model the large scale structure), and the \astrid\ simulation (to model the halo-quasar luminosity relation and the proximity effect fitting uncertainties).

 There are a number of simulations in the \uchuu\ suite \footnote{\url{skiesanduniverses.org}}, and we use the largest, \uchuu\ itself. This consists of a cubic periodic 
 volume of side-length 2000 \chmpc, containing $12800^{3}$ dark matter particles. We work with the $z=3$ output exclusively, so our simulation will not include any lightcone effects. The mass resolution is  $2.3\times10^{8}$\hmsun\ per particle, and the Rockstar halo finder (\citealt{rockstar}) was used to produce a dataset of $5.8\times 10^{6}$ halos. In order to match the space density of
 the top 1000 most luminous halos we have used from \astrid, we restrict the \uchuu\ dataset to the top 512,000 halos by luminosity (see below for assignment of luminosity). This is approximately equal to those above a lower mass limit of $4.6\times10^{12}\msun$ or 20,000 particles.
 
 We use the relationship shown in 
Figure \ref{bhproperties_mhalo} to assign $r_{\rm eq}$ values to
each halo in \uchuu:
\begin{equation}
\label{eq:halo_mass_req}
\log_{10}\bigl(r_{\mathrm{eq}}\bigr)
\;=\;
-8.6
\;+\;
0.70\,\log_{10}\bigl(M_{\mathrm{halo}}\bigr)
\;+\;
0.25\,\epsilon,
\end{equation}
where \(\epsilon \sim \mathcal{N}(0,1)\) is a standard normal deviate capturing 
the intrinsic scatter.
 The $r_{\rm eq}$ in each case is related to the quasar luminosity according to
Equation \ref{req_eqn}. We then use the inverse square law 
attenuated by an attenuation length to assign the quasar
radiation to the positions of all the other quasars:

\begin{equation}
\Gamma_{j} = f_{d}\sum_{i\neq j}\frac{r^{2}_{\rm eq}}{r^{2}_{ij}+r^{2}_{p}}.
\label{invsq}
\end{equation}

Here the units of $\Gamma$ are $10^{-12}$ s$^{-1}$, which were the units used to define $r_{\rm eq}$
in Figure \ref{reqindiv}. The distance $r_{ij}$  in Equation \ref{invsq} is the separation between two quasars $i$ and $j$. The quantity $r_{p}$
is a Plummer-like softening radius (\citealt{plummer1911}) used to avoid the radiation 
intensity blowing up at small radii. We use
$r_{p}=1.0$ \hmpc but find that our final
results are insensitive to variations of
$r_{p}$ over the range 0.1 to 10 \hmpc . We assume that quasars have a lifetime at least as long as the light travel time across half the box (1.2 Gyr), so that all quasars within that distance contribute. This is an assumption which could be varied (see e.g., \citealt{croft04})  using the actual quasar lifetimes from hydrodynamic simulations 
(\citealt{zhou24}) but we leave this to future work. We do however include a factor $f_{d}$ in Equation \ref{invsq} which is the mean fraction of the instantaneous quasar light which reaches a point in the IGM. This factor is meant to crudely model effects including the quasar duty cycle, and the fraction of sightlines unobscured by the host galaxy or gas close to the quasar. We set the value of this parameter to $f_{d}=0.01$, which yields a mean value
of $\Gamma=0.8\times 10^{-12}$ s$^{-1}$, consistent with the observed value at $z=3$ (\citealt{fg20}). 

Now that the radiation field is varying spatially, the actual value of $r_{\rm eq}$ [which was
computed assuming $\Gamma=1.0$ ($\times 10^{-12}$ s$^{-1}$)]  will be modified in response to the
local UVBG in this way:

\begin{equation}
r'_{\rm eq,i} = r_{\rm eq,i}/\sqrt{\Gamma_{i}},
\end{equation}
where $r'_{\rm eq}$ is the new value. 

We then determine the accuracy of the halo model fitting to the proximity effect to
 measure the local value of $\Gamma$ for each quasar.
 We do this using the results from \astrid, specifically the rms error on $r_{\rm eq}$ for fits to individual quasars as a
function of the true $r_{\rm eq}$ values shown in Figure \ref{reqindiv}. By plotting the pdf of fit values, we 
find that the error in 
the measured $r_{\rm eq}\prime$ for each quasar is Normally distributed. We therefore model the effect of the fitting process
by adding a random fitting error  $r'_{{\rm err},i}$ so that
$r'_{\rm eq,i}\prime_{fitted} = r'_{\rm eq,i} + r'_{{\rm err},i}$. The values of $r'_{{\rm err},i}$ are Normally distributed with an
rms given by $\sigma_{{\rm err},i}=0.25 r'_{\rm eq,i}+2.03 $, derived from the results shown in Figure \ref{erroronreq}.

\begin{figure*}
    \begin{centering}
    \includegraphics[width=0.7\linewidth]{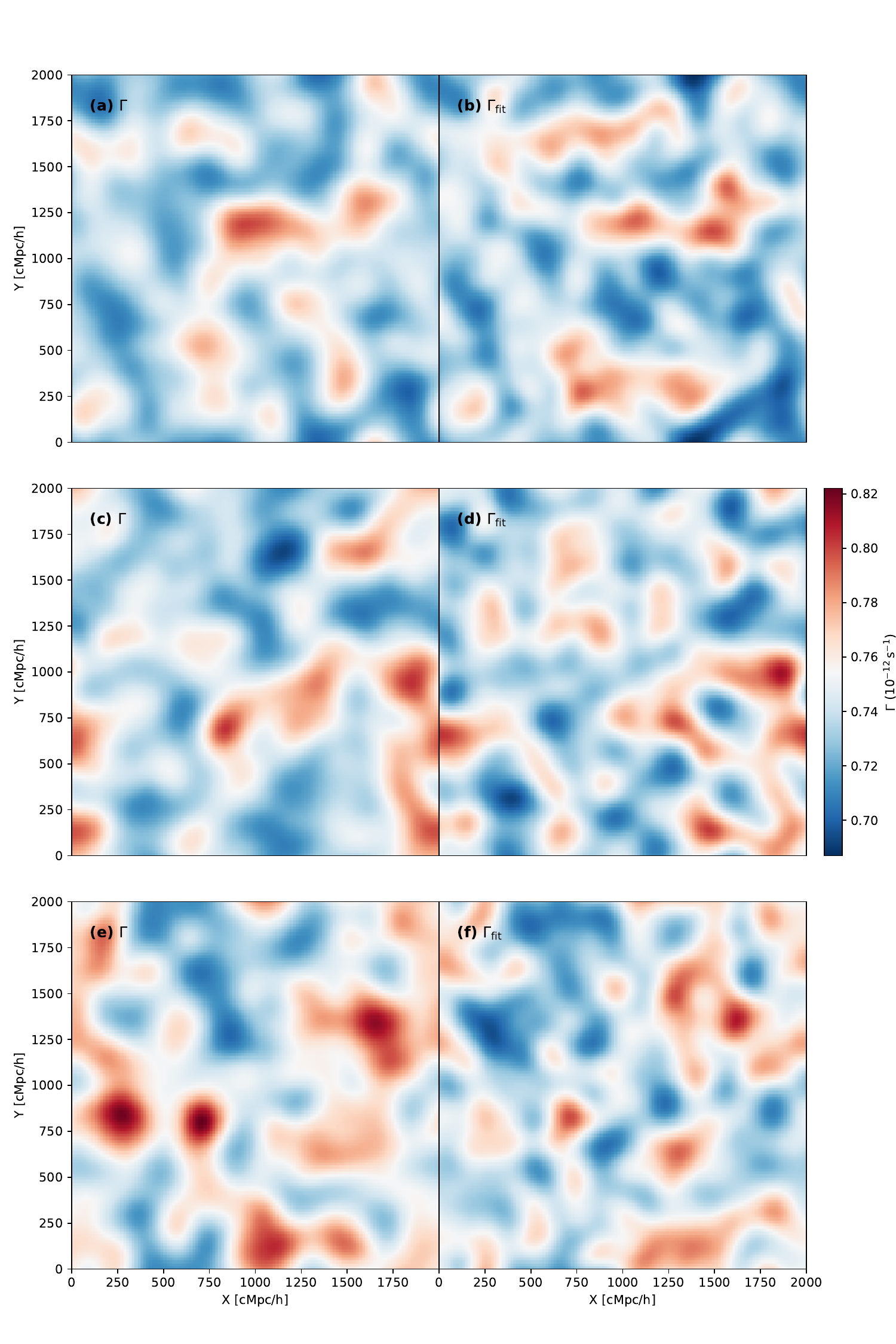}
    \caption{Slices through the ultraviolet 
    background radiation field in the \uchuu\
    simulation at redshift $z=3$. The left panels show the true field and the right panels show the field fitted from the proximity effect. The slices are in the x-y plane, at  $z$-positions  of z=333, 1000  and 1667 \chmpc.
    }
    \label{slices}
    \end{centering}
\end{figure*}

The values of the UVBG intensity that result from these proximity effect
fits are 
\begin{equation}
\Gamma_{i, {\rm fitted}}=\frac {\Gamma_{i} r'^{2}_{\rm eq,i}}{r'^{2}_{\rm eq,i,{fitted}}+r^{2}_{p}},
\end{equation}

where  we again use the
Plummer-like softening $r_{p}$ (same values as in Equation \ref{invsq}) to avoid large errors in the radiation field due to small uncertainties in the quasar positions.
These values of  $\Gamma_{i} $  and $\Gamma_{i, {\rm fitted}}$ represent the true and the proximity effect-fitted values of the ionizing background radiation
at the positions if the quasars. These positions are
quite sparse, and so it is likely that the best way to analyse the spatial fluctuations in $\Gamma$ would be to
compute the autocorrelation function of the $\Gamma$ values as a function of scale. Here though, we are 
interested in visualizing and quantifying the continuous
radiation field. We therefore leave analysis of the 
autocorrelation function to future work and instead
assign the individual $\Gamma$ values to a cartesian grid. We do this using a Triangular Shaped Clouds (\citealt{hockneyandeastwood}) assignment scheme, using the redshift-space coordinates of the quasars. The grid we use to cover the \uchuu\ volume has $128^{3}$ cells.

In Figure \ref{slices} we show slices through the radiation field, showing the $x-y$ plane at 3 different $z$ positions. We have smoothed the 
field with a Gaussian distribution of $\sigma=75 \chmpc$
because the unsmoothed fields are visually
noisy (we analyze the unsmoothed fields using the power spectrum below). In order to make the color scales comparable to make visual comparisons easier, we scale the fitted $\Gamma_{i}$ field so that the $rms$ fluctuations about the mean are the same as the true field. Without this, the noise fluctuations in the  fitted field make it more difficult to pick out structures, even with smoothing. We can see from Figure \ref{slices} that the real and proximity effect-fitted fields are broadly similar, as the true (left) and fitted (right) field pairs can be matched. Individual
structures are not well reproduced, but overall it seems as though the proximity effect can be used to produce a somewhat noisy (we quantify this below) map of the intergalactic radiation field. The rms fluctuations on $\sim 100 \chmpc$ scales are of order 10\% of the mean, comparable to those in the density field (as we see below).

\begin{figure}
    \centering
    \includegraphics[width=\linewidth]{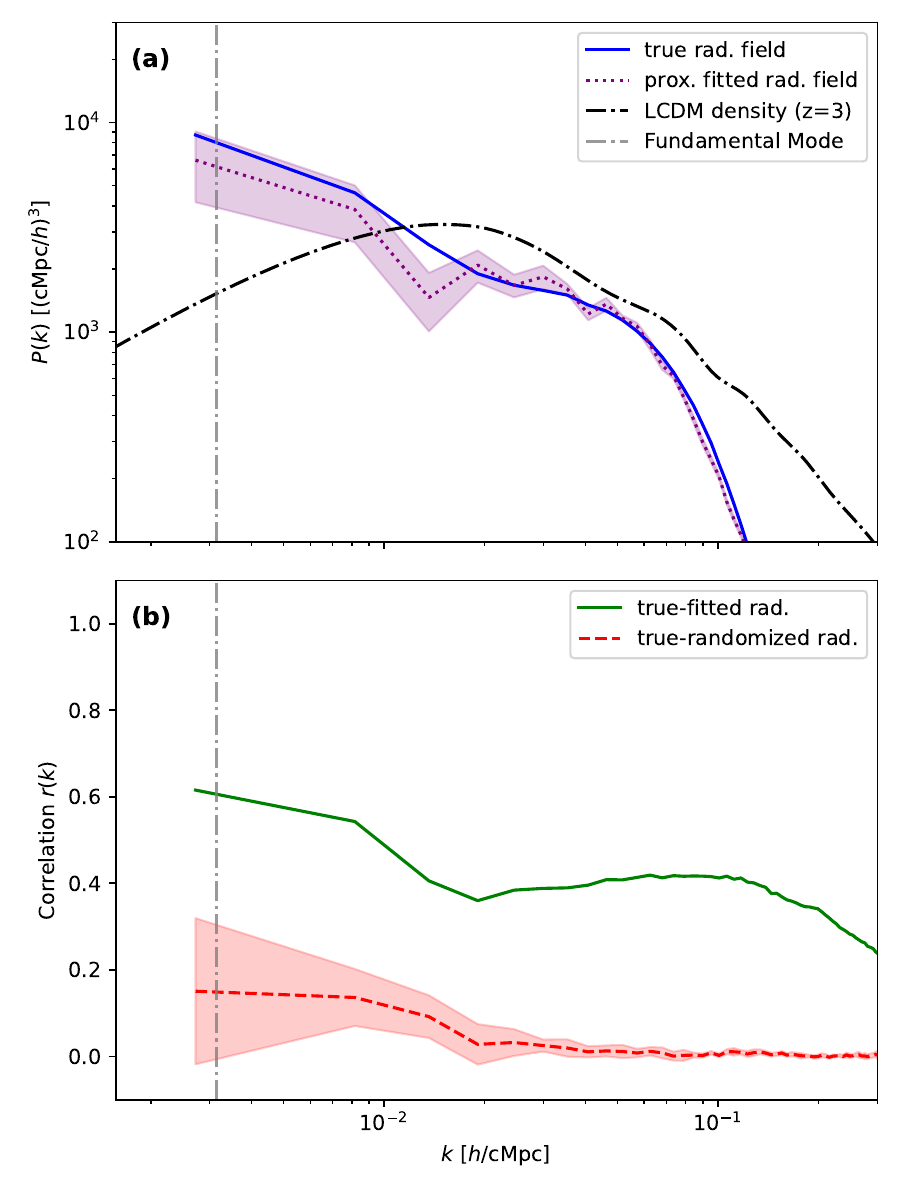}
    \caption{(a) Power spectrum of the UVBG radiation field in the \uchuu\ simulation at redshift $z=3$, as well as the power spectrum of the UVBG in \uchuu\ estimated by fitting the proximity effect. A position-randomized ("noise") field was subtracted from the proximity effect derived field prior to plotting (see text). The shaded region shows the error on the mean from subtraction of those 10 random realizations. We also show the linear power spectrum of the matter density. (b) The correlation coefficient of individual Fourier modes in the true radiation field and the one fitted from the proximity effect. We also show the correlation coefficient for the true field and the randomized field along with the standard deviation among 10 random realizations as a shaded area.
    }
    \label{pkcorr}
\end{figure}

To quantify the noise level in the reconstructed map, we measured the correlation coefficient $r(k)$ between individual Fourier modes of the true and reconstructed fields. We show this as a function of wavenumber $k$ in the bottom panel of Figure \ref{pkcorr}.  The fields are correlated at the $r\sim 0.4-0.6$ level for wavenumbers from the fundamental mode of the volume (plotted) to $k \sim 0.1$ \invchmpc. The Nyquist frequency of the grid is 0.2 \invchmpc, although the sparsity of the quasar sample and the TSC assignment both combine to limit the smallest scales that can be probed. 

In the lower panel of Figure \ref{pkcorr} we also show the cross-correlation  of Fourier modes between the true field and a randomized field. We construct the randomized field by replacing the cartesian coordinates of the quasars in \uchuu\ with random coordinates drawn from a uniform distribution. We then generate the radiation field using the inverse square law as before and then  the $\Gamma $ values using Equation \ref{invsq}. This randomized field embodies the shot noise component of the proximity effect fitting procedure but not the large-scale correlations. We generate 10 different realizations with different random seeds. The mean correlation coefficient of these realizations with the true field is shown in the figure, along with the 1 $\sigma$ standard deviation among the 10 realizations. We can see that $r(k)$ is much lower than the correlation with the true field, being consistent with zero at the 1 $\sigma$ level for most of the $k$ range.

The clustering of the radiation field can be measured using the
power spectrum, and we show this in the top panel of Figure \ref{pkcorr}, alongside the linear power spectrum of the LCDM density field at $z=3$.
In the same manner as is done for the density fluctuations (where the relative overdensity $\delta_{\rho}=(\rho/\rhobar)-1$ is calculated), we compute the
relative overintensity  $\delta_{\Gamma}=(\Gamma/\overline{\Gamma})-1$ (using the photoionization rate $\Gamma$ to quantify the radiation intensity). $P(k)$ for the
relative overintensity is shown for the true field, but 
because the reconstructed field from the proximity 
effect is noisy, we need to subtract the noise contribution to $P(k)$
in order to compare to $P(k)$ for the true radiation field. Although 
the shot noise comes from Poisson-like component, it does not have a power spectrum which is strictly a constant, because of the
inverse square profile of radiation around sources and the fact that 
they are of widely different luminosities.
We therefore use the
randomized realizations that we already used in our null test of the
cross-correlation of Fourier modes above for the noise 
subtraction. We compute $P(k)$ for the
10 realizations of the randomized field and subtract their mean
from $P(k)$ for the proximity-fitted field. We plot the noise-corrected
proximity field power spectrum and the error on the mean $P(k)$ from the 10 different realizations. We can see that the noise-corrected power spectrum agrees quite well with the power spectrum of the true radiation field.

The power spectrum of the intergalactic radiation field fluctuations 
in Figure \ref{pkcorr}
is similar in amplitude to that of the matter power spectrum. There is a rapid dropoff
on small scales $k>0.05$ h cMpc$^{-1}$, which is likely due to the sparsity of sources and the resolution of the radiation field grid. On large scales, $P(k)$ has an
approximately power-law shape, with a power-law slope $n\sim -0.7$. The radiation
power spectrum has a larger amplitude than the matter power spectrum on scales $k< 0.01$ h cMpc$^{-1}$.
In detail though this may depend on aspects of the simulated radiation field and sources that have not been varied, including quasar lifetime, beaming and obscuration, and radiative transfer through the intergalactic medium. These proximity
effect measurements may be useful to probe these and perhaps the cosmological model. We leave this for future work.

\section{Summary and discussion}

\label{sumdisc}

\subsection{Summary}

We have carried out modeling of the quasar proximity effect using a cosmological hydrodynamic simulation and an analytic model based on the clustering and structure of dark matter halos in a CDM-dominated 
universe. Our emphasis is on developing modeling techniques and analysis methods that could be 
applied to large samples of quasars from current and future spectroscopic surveys. The main findings
from our study can be summarized as follows.

(a) Our halo model for the proximity effect is related to the halo model used to study
galaxy clustering (e.g., \cite{cooray02}). As such it takes into 
account the increased overdensity near quasars and so provides
an unbiased estimate of the effect of ionizing radiation on that material. The model includes redshift distortions
and local ionizing radiation from the quasar and the ultra-violet 
background, both in  the optically thin approximation.

(b) We have extracted a sample of quasars from the \astrid\ simulation at $z=3$ with 
bolometric luminosities from $10^{45}\ergs$ to $10^{47}\ergs$. We find relationships
 between host halo mass and quasar luminosity  as well as host halo mass and black hole mass,
 with both having substantial scatter (0.1-0.4 dex).
The quasar-matter cross-correlation can be fit well using the halo  model, with
brightest quasars in \astrid\ ($L\sim 10^{47}$ ergs$^{-1}$) having a linear bias of $b_{\rm q}=4$.

(c) We have tested our proximity effect halo model using the \lya\ forest from 
\astrid, finding that we can fit for the $b_{\rm q}$ and $r_{\rm eq}$
parameters. The radius of equality between local and background ionizing radiation, $r_{\rm eq}$, is a measure which is directly related to the observable, the
radial variation of transmitted flux in \lya\ forest spectra.
The confidence contours of the fit are surprisingly independent between the two parameters. We can fit $r_{\rm eq}$ with 25-50\% accuracy for individual quasars, depending on their
luminosity. We also find that we could constrain the host halo mass
from the fitted $b_{\rm q}$ parameter at the $\sim$ 10\% level for averages over samples of 200 quasars.

(d) Although quasars are sparse, we have seen that with the space density of current and upcoming surveys we should be able to use the proximity effect to map
out the cosmic radiation field spatially.  From a large dark matter simulation and measurements from the hydrodynamic run, we simulated recovery of the radiation field on 3D grid covering
2 \chgpc\ in diameter. We find that the $P(k)$ of the radiation is measurable on scales from
$k\sim 0.001$ to $\sim0.1 \invchmpc$. It has similar amplitude to the matter $P(k)$,
although future work will be needed to predict its dependence on source lifetime, beaming, and lightcone effects.

\subsection{Discussion}
\label{discussion}

The fact that the halo model (at least in simulation tests) is unbiased with
respect to the quasar radiation and density fluctuations is a positive
aspect. The effect of overdense regions close to quasars partially cancelling effect of local ionizing radiation is something that has been studied extensively (e.g., \cite{guimar07}). We
hope that the halo model can give some extra clarity so that the proximity effect
can be used more widely as a probe of the high redshift Universe.
At the very least new  large surveys of quasars will be able to use
the effect to yield corroborating information about the current model
of structure formation and also quasars and their luminosity output.
Mapping the intergalactic radiation field is a new aspect which will 
require many sources. The preliminary test in this paper uses a space density of quasars which is similar to but still higher than that of the current DESI survey.
 Further work is needed to investigate the signal to noise of the map as a function of source space density. We anticipate that using fainter quasars than in DESI (see e.g., \citealt{martini25}) should not change the map drastically because the most luminous sources that are in DESI will have well-measured proximity effects and will dominate the map.

There are many complications and uncertainties that have not been included in our
analysis so far. One of the most important is that the model can measure the radius
$r_{\rm eq}$ quite well, but in order to determine the photoionization rate $\Gamma$ we also need to know the quasar ionizing radiation output. Basing this on the 
observed quasar magnitude will lead to errors due to quasar luminosities varying 
with time. For example, the equilibration time of the photoionized medium
is $\sim 10^{4}$ yrs, but quasars can vary significantly on timescales
less than this (see e.g., \citealt{wagner95} for observations or \citealt{zhou24} for
simulations). In our modelling we have also used the optically thin approximation 
(albeit attentuated by a uniform mean free path). The presence of dense gas close to quasars could change how appropriate this is, at least for some sightlines
(\citealt{ni20}).

In addition to these uncertainties, future work could also usefully investigate the following: 

(a) how the S/N of the reconstructed radiation field depends on the number density of quasars used, and also on realistic levels of noise in the observed quasar luminosities and in the \lya\ forest spectra themselves. 

(b) How the background radiation and its recovery through the proximity effect depend on models for source emission and evolution: e.g., beaming and light cone effects. 

(c) The photon mean free path could be  varied
to see how this affects the large-scale clustering of the radiation field.  

(d) Redshifts beyond 
the $z=3$ we have used could be studied, including those earlier times when the optically thin
approximation may no longer be appropriate.

(e) Observations of quasar spectra also have their own systematic 
sources of uncertainty. These include redshift uncertainties on the position of the quasar itself, which can be severe enough to erase all evidence of the proximity effect (e.g., \citealt{kirkman08}). Fitting the quasar unabsorbed continuum close to the \lya\ line is also potentially difficult (\citealt{liu21}).

The \lya\ forest itself can tell us about the UVBG because it affects both the mean transmitted
flux and clustering in the forest. Early work such as \cite{rauch97} showed
how the flux pdf could constrain the mean intensity, assuming accurate modeling of the
IGM on the $\sim 100 \hkpc$ scales relevant to the pressure smoothing in the forest. Comparison with proximity effect measurements can help to constrain the small scale gas physics as well as the sources. In terms of the forest clustering (\citealt{mcquinn11,pontzen14}), there could be role that proximity 
effect measurements could play when constraining the radiation fluctuations in the context of
nuisance parameters for  BAO measurements (\citealt{gontcho14,pontzen14}). In addition, absorption by different species of metal lines is sensitive to UV background spatial fluctuations (\citealt{morrison21}) and could in principle be used to trace them out and compare to inferences from the proximity effect.

Competitive constraints on the cosmological model are not likely with the
proximity effect measurements on their own. It could in principle be interesting, 
because it allows one to compare a direct measurement of quasar luminosity (from observed magnitude)
with an indirect one (proximity effect), and so constrain the distance scale. This was first pointed out by 
\cite{phillipps02}, but even then the many likely uncertainties seemed to make this very difficult.
Now the precision required for a competitive measurement is even higher. We can see this by
imagining that we have a very large sample of proximity effect measurements at different redshifts, say at $z=2$ and $z=3$. If so, we would be measuring the $r_{\rm eq}$ distance at different redshifts
and so effectively determining the difference in the Hubble parameter
$H(z)$ between $z=2$ and $z=3$.  Using the relation appropriate for a flat 
cosmology, $H(z)=H_0 \sqrt(\Omega_{m} (1+z)^3 +\Omega_{\Lambda})$ 
we find that the $H(z=3)/H(z=2)$ changes by -0.074 \% for every increase in
$\Omega_{\Lambda}$ by 1\%. Assuming we have 40\% error on $r_{\rm eq}$ from each quasar we would 
need 30 million quasars just to measure $\Omega_{\Lambda}$   to 10\%. Not only is
this not feasible, but the systematic uncertainties detailed above would likely outweigh the statistical errors (not to mention the fact that for a useful measurement we would need an independent measurement of the ionizing background intensity as a function of redshift).

Rather than constraining the distance scale, it is probably better to think of  large proximity effect surveys as giving information on the sources of radiation. The large scales seen in the radiation field power spectrum  (for example the  high amplitude beyond the
matter-radiation equality peak in the matter P(k))
are certainly interesting and may be useful cosmologically. We first need to see
how they are affected by finite lifetime effects and beaming, which are
likely to  "chop up" the field (see e.g., Figure 5 of \citealt{croft04}). 
There are also likely to be better ways to analyse the radiation fluctuations, 
as assigning the individual $\Gamma$ values to a grid is not really appropriate
for such sparse sources. Well-specified techniques for interpolating sparse data into a volume-filling map could be borrowed from other areas (e.g., \citealt{cisewski14},
\citealt{pichon01}, \citealt{ozbek16}, \citealt{malicke22}). These could also be used to investigate  biases in the field reconstruction associated with the fact that the proximity effect probes (the quasars) are also the sources and so are clustered in regions of space with a high radiation intensity.

%\begin{equation}
%  \frac{\delta(R_{3-2})/R_{3-2}}{\delta \Omega_\Lambda/\Omega_\Lambda}
%\simeq  -0.074,
%\end{equation}
%where $R_{2-3}=H(z=3)/H(z=2)$.

The proximity effect is the result of a fascinating interaction between the 
brightest sources of radiation in the Universe and the diffuse, low density
intergalactic medium. It is directly detectable in individual spectra over length scales of ten or more (proper) Megaparsecs, much further than any other non-gravitational effect such as  galactic winds or feedback. Huge quasars surveys  produce data that  has not yet been analyzed with this effect in mind, and the
statistical power that will be available should allow us to uncover new science in the intergalactic radiation field.

\section*{Acknowledgements}

\texttt{ASTRID} was run on the Frontera facility at the Texas Advanced Computing Center.
TDM acknowledges additional support from  NSF ACI-1614853, NSF AST-1616168, NASA ATP 19-ATP19-0084, and NASA ATP 80NSSC20K0519. SB acknowledges funding supported by NASA-80NSSC22K1897.
The Instituto de Astrofisica de Andalucia (IAA-CSIC), Centro de Supercomputacion de Galicia (CESGA) and the Spanish academic and research network (RedIRIS) in Spain host \uchuu\ DR1, DR2 and DR3 in the Skies \& Universes site for cosmological simulations. The \uchuu\ simulations were carried out on Aterui II supercomputer at Center for Computational Astrophysics, CfCA, of National Astronomical Observatory of Japan, and the K computer at the RIKEN Advanced Institute for Computational Science. The \uchuu\ Data Releases effort made use of the \texttt{skun@IAA\textunderscore RedIRIS} and \texttt{skun6@IAA} computer facilities managed by the IAA-CSIC in Spain (MICINN EU-Feder grant EQC2018-004366-P).

\bibliographystyle{mnras}
\bibliography{ref}

\end{document}